\documentclass[10pt,twocolumn]{article}

\usepackage[utf8]{inputenc}
\usepackage[T1]{fontenc}
\usepackage[margin=0.78in]{geometry}
\usepackage{amsmath,amssymb}
\usepackage{array}
\usepackage{booktabs}
\usepackage{tabularx}
\usepackage{flushend}
\usepackage{xcolor}
\usepackage[hyphens]{url}
\usepackage[colorlinks=true,linkcolor=black,citecolor=blue!60!black,urlcolor=black]{hyperref}

\makeatletter
\renewcommand\section{\@startsection{section}{1}{\z@}%
  {-3.2ex \@plus -1ex \@minus -.2ex}%
  {1.8ex \@plus .2ex}%
  {\normalfont\Large\bfseries\raggedright}}
\renewcommand\subsection{\@startsection{subsection}{2}{\z@}%
  {-2.8ex \@plus -0.8ex \@minus -.2ex}%
  {1.2ex \@plus .2ex}%
  {\normalfont\large\bfseries\raggedright}}
\makeatother

\title{\Large\textbf{Epistemic Guard: Out-of-Process MCP Receipt Proxying,\\
Post-Turn Prose Supervision, and Provenance-Gated Memory\\
for Tool-Using Autonomous Agents}}

\author{
  Sean Maxwell\\
  Roll Solo, Melbourne, VIC, Australia\\
  \texttt{sean.maxwell@rollsolo.app} \quad
  \url{https://orcid.org/0009-0004-0133-0463}
}

\date{November 2026}

\begin{document}

\twocolumn[
\begin{@twocolumnfalse}
\vspace{-1.8em}
\maketitle
\vspace{-1.2em}
\begin{center}
\begin{minipage}{0.94\textwidth}
\small
\noindent\textbf{Abstract---}Language-model agents using the Model Context Protocol (MCP) face three runtime verification gaps: (i) \emph{action fabrication}, where post-tool prose narrates side effects that failed, timed out, or never ran; (ii) \emph{provenance collapse}, where unverified model prose or web text enters durable memory alongside executed tool outputs; and (iii) \emph{syndicated echo amplification}, where mirrored web pages satisfy multi-source corroboration from one origin. Existing MCP firewalls inspect wire calls before execution, while LLM-judge guardrails require multi-gigabyte VRAM without binding prose to receipts. We present \texttt{epistemic-guard} (v0.5.2), an open-source Python library and out-of-process \texttt{stdio} MCP receipt proxy (\href{https://github.com/AlienatedAliens/epistemic-guard/blob/v0.5.2/results/table1_architecture.json\#L30}{15} modules, \href{https://github.com/AlienatedAliens/epistemic-guard/blob/v0.5.2/results/table3_heldout_supervision.json\#L84}{0}\,MiB GPU VRAM). \texttt{epistemic-guard} isolates Ed25519 signing keys out of process, writes RFC\,6962 Merkle logs with \texttt{AER-1} receipt translation, checks post-turn prose on CPU with SHA-256-verified OpenVINO INT8 encoders (\href{https://github.com/AlienatedAliens/epistemic-guard/blob/v0.5.2/results/table1_architecture.json\#L16}{4} models, \href{https://github.com/AlienatedAliens/epistemic-guard/blob/v0.5.2/results/table1_architecture.json\#L34}{279.48}\,MB total; \href{https://github.com/AlienatedAliens/epistemic-guard/blob/v0.5.2/results/table1_architecture.json\#L19}{82.67}\,MB on the supervision path), gates SQLite writes across \href{https://github.com/AlienatedAliens/epistemic-guard/blob/v0.5.2/results/table1_architecture.json\#L28}{7} provenance domains, and collapses verbatim and lightly edited wire-syndicated RAG mirrors. On conformance suites, \texttt{epistemic-guard} rejects \href{https://github.com/AlienatedAliens/epistemic-guard/blob/v0.5.2/results/table2_conformance.json\#L44}{100.0\%} of tampered receipts (\href{https://github.com/AlienatedAliens/epistemic-guard/blob/v0.5.2/results/table2_conformance.json\#L43}{200} Merkle leaves), passes \href{https://github.com/AlienatedAliens/epistemic-guard/blob/v0.5.2/results/table2_conformance.json\#L18}{58}/\href{https://github.com/AlienatedAliens/epistemic-guard/blob/v0.5.2/results/table2_conformance.json\#L21}{58} official \texttt{AER-1} ingestion vectors and \href{https://github.com/AlienatedAliens/epistemic-guard/blob/v0.5.2/results/table2_conformance.json\#L22}{50}/\href{https://github.com/AlienatedAliens/epistemic-guard/blob/v0.5.2/results/table2_conformance.json\#L14}{50} reference-verified \texttt{AER-1} exports, verifies \href{https://github.com/AlienatedAliens/epistemic-guard/blob/v0.5.2/results/table2_conformance.json\#L39}{100.0\%} of \href{https://github.com/AlienatedAliens/epistemic-guard/blob/v0.5.2/results/table2_conformance.json\#L34}{40} fault-injected wire calls and \href{https://github.com/AlienatedAliens/epistemic-guard/blob/v0.5.2/results/table2_conformance.json\#L5}{100} concurrent appends with \href{https://github.com/AlienatedAliens/epistemic-guard/blob/v0.5.2/results/table2_conformance.json\#L28}{0} key leaks, and blocks \href{https://github.com/AlienatedAliens/epistemic-guard/blob/v0.5.2/results/table2_conformance.json\#L60}{100.0\%} of \href{https://github.com/AlienatedAliens/epistemic-guard/blob/v0.5.2/results/table2_conformance.json\#L58}{100} unauthorized memory writes. Across $N=\href{https://github.com/AlienatedAliens/epistemic-guard/blob/v0.5.2/results/table3_heldout_supervision.json\#L175}{250}$ frozen-benchmark traces, \texttt{epistemic-guard} reaches \href{https://github.com/AlienatedAliens/epistemic-guard/blob/v0.5.2/results/table3_heldout_supervision.json\#L71}{93.2\%} accuracy at \href{https://github.com/AlienatedAliens/epistemic-guard/blob/v0.5.2/results/table3_heldout_supervision.json\#L86}{17.9}\,ms warm median CPU latency (\href{https://github.com/AlienatedAliens/epistemic-guard/blob/v0.5.2/results/table3_heldout_supervision.json\#L97}{8.62}\,ms single-clause; \href{https://github.com/AlienatedAliens/epistemic-guard/blob/v0.5.2/results/table3_heldout_supervision.json\#L78}{921.8}\,ms cold-start load), beating 12B--14B GPU judges by \href{https://github.com/AlienatedAliens/epistemic-guard/blob/v0.5.2/results/table3_heldout_supervision.json\#L24}{6.4}--\href{https://github.com/AlienatedAliens/epistemic-guard/blob/v0.5.2/results/table3_heldout_supervision.json\#L25}{6.8} points ($p=\href{https://github.com/AlienatedAliens/epistemic-guard/blob/v0.5.2/results/table3_heldout_supervision.json\#L36}{0.0259}$, $p=\href{https://github.com/AlienatedAliens/epistemic-guard/blob/v0.5.2/results/table3_heldout_supervision.json\#L39}{0.0095}$) at \href{https://github.com/AlienatedAliens/epistemic-guard/blob/v0.5.2/results/table3_heldout_supervision.json\#L41}{4.26}--\href{https://github.com/AlienatedAliens/epistemic-guard/blob/v0.5.2/results/table3_heldout_supervision.json\#L40}{4.5}$\times$ lower latency; on $N=\href{https://github.com/AlienatedAliens/epistemic-guard/blob/v0.5.2/results/table3_heldout_supervision.json\#L136}{60}$ live \texttt{mcp-proxy} traces, it catches \href{https://github.com/AlienatedAliens/epistemic-guard/blob/v0.5.2/results/table3_heldout_supervision.json\#L121}{96.15\%} of hazards at \href{https://github.com/AlienatedAliens/epistemic-guard/blob/v0.5.2/results/table3_heldout_supervision.json\#L108}{76.67\%} accuracy (\href{https://github.com/AlienatedAliens/epistemic-guard/blob/v0.5.2/results/table3_heldout_supervision.json\#L116}{38.24\%} FPR; $+\href{https://github.com/AlienatedAliens/epistemic-guard/blob/v0.5.2/results/table3_heldout_supervision.json\#L107}{18.34}$ points over Stage-1-only, $p=\href{https://github.com/AlienatedAliens/epistemic-guard/blob/v0.5.2/results/table3_heldout_supervision.json\#L133}{0.001}$) in \href{https://github.com/AlienatedAliens/epistemic-guard/blob/v0.5.2/results/table3_heldout_supervision.json\#L117}{9.16}\,ms p50.

\smallskip
{\footnotesize\raggedright
\noindent\textbf{Artifacts:}
\mbox{Code: \url{https://github.com/AlienatedAliens/epistemic-guard}} $\cdot$
\mbox{PyPI: \url{https://pypi.org/project/epistemic-guard/}} $\cdot$
\mbox{INT8 weights: \url{https://huggingface.co/RollSolo/epistemic-guard-int8}} $\cdot$
\mbox{Benchmarks: \url{https://huggingface.co/datasets/RollSolo/epistemic-guard-benchmarks}}\par}
\end{minipage}
\end{center}
\vspace{0.8em}
\hrule height 0.4pt
\vspace{1.0em}
\end{@twocolumnfalse}
]

\section{Introduction}
\label{sec:intro}

Software engineering and operations agents invoke external tools over JSON-RPC transports such as the Model Context Protocol (MCP)~\cite{mcp2025spec}. MCP standardizes wire transport, not post-call truthfulness: what happened on the wire can diverge from what the language model tells the user or writes to persistent state. In live agent loops, tool calls return HTTP errors, time out, fail schema checks, or never run at all while the model claims in first-person prose that the action succeeded~\cite{maxwell2026evasion}. When downstream workflows or memory stores ingest those unverified claims, single-turn hallucinations persist across sessions~\cite{greshake2023indirect,maxwell2026poisoning}.

From a runtime verification perspective~\cite{leucker2009brief,bartocci2018lectures}, closing this gap without trusting the agent's own prompt context or reserving GPU memory requires three checks across each turn:
\begin{enumerate}
  \item \textbf{Out-of-Process Execution Receipts:} Every MCP tool call and response must be signed outside the agent process, separating pre-execution parameter rejection (\texttt{executed=false}) from runtime tool failure (\texttt{executed=true, ok=false}).
  \item \textbf{Receipt-Bound Prose Supervision:} Before an assistant reply reaches the user or a downstream workflow, state-mutating claims in natural language must match verified execution receipts and arguments from the current turn.
  \item \textbf{Provenance-Gated Storage and De-Syndicated RAG:} Persistent memory stores must reject unsigned or narration-sourced writes at the database boundary, and retrieval-augmented generation (RAG) pipelines must collapse mirrored web copies before counting independent sources on high-hazard queries.
\end{enumerate}

These failure modes correspond to three entries in the OWASP Top 10 for Agentic Applications~\cite{owasp2026agentic}: Tool Misuse and Unverified Action Claims (\texttt{ASI02}), Memory and Context Poisoning (\texttt{ASI06}), and Cascading Inter-Component Hallucinations (\texttt{ASI08}). Existing open-source tools cover separate parts of this path. Transport firewalls such as Pipelock~\cite{pipelock2026} scan MCP egress traffic and emit mediator-signed receipts, and runtimes such as NabaOS~\cite{nabaos2026} enforce Ed25519-signed policy constitutions, but neither checks post-turn prose against tool receipts nor deduplicates syndicated RAG sources before memory writes. Dialogue guardrails such as NeMo Guardrails~\cite{rebedea2023nemo}, Llama Guard~\cite{inan2023llamaguard}, and Guardrails AI~\cite{guardrailsai2024} filter text with dialogue rules or LLM judges that consume \href{https://github.com/AlienatedAliens/epistemic-guard/blob/v0.5.2/results/table3_heldout_supervision.json\#L166}{8,844}--\href{https://github.com/AlienatedAliens/epistemic-guard/blob/v0.5.2/results/table3_heldout_supervision.json\#L182}{12,214}\,MiB of GPU VRAM and do not verify cryptographic tool receipts.

This paper describes the architecture, security invariants, and empirical evaluation of \texttt{epistemic-guard} (v0.5.2), a Python runtime verification library and MCP proxy~\cite{maxwell2026seed,maxwell2026fivelayer}. Its main components are:
\begin{itemize}
  \item \textbf{Out-of-Process MCP Receipt Proxy (\texttt{mcp-proxy}):} A \texttt{stdio} JSON-RPC\,2.0 proxy that strips signing secrets from child server environments, classifies \href{https://github.com/AlienatedAliens/epistemic-guard/blob/v0.5.2/results/table2_conformance.json\#L31}{4} execution and error states, appends Ed25519-signed receipts to an OS-locked JSONL Merkle log with public-key pinning, and translates bidirectionally to the Agent Execution Receipt (\texttt{AER-1}) format~\cite{zambo2026aer1}.
  \item \textbf{CPU INT8 Receipt Supervision:} A hybrid symbolic--neural supervisor (\texttt{ReceiptSupervisor}) using the fine-tuned \texttt{action\_claim} OpenVINO INT8 cross-encoder (\href{https://github.com/AlienatedAliens/epistemic-guard/blob/v0.5.2/results/table1_architecture.json\#L19}{82.67}\,MB of the \href{https://github.com/AlienatedAliens/epistemic-guard/blob/v0.5.2/results/table1_architecture.json\#L34}{279.48}\,MB \href{https://github.com/AlienatedAliens/epistemic-guard/blob/v0.5.2/results/table1_architecture.json\#L16}{4}-model bundle; \href{https://github.com/AlienatedAliens/epistemic-guard/blob/v0.5.2/results/table3_heldout_supervision.json\#L84}{0}\,MiB GPU VRAM) plus Ed25519 receipt signature and argument-target matching, running in \href{https://github.com/AlienatedAliens/epistemic-guard/blob/v0.5.2/results/table3_heldout_supervision.json\#L86}{17.9}\,ms median turn CPU latency.
  \item \textbf{Provenance-Gated SQLite Store \& RAG Mirror Filter:} Domain-separated HKDF-SHA256~\cite{krawczyk2010rfc5869} + Ed25519~\cite{josefsson2017rfc8032} write-gating with SQLite immutability triggers, paired with word-shingle Jaccard and SimHash clustering~\cite{broder1997resemblance,charikar2002similarity} across a \href{https://github.com/AlienatedAliens/epistemic-guard/blob/v0.5.2/results/table1_architecture.json\#L26}{12}-cell hazard-by-proof-standard policy table.
  \item \textbf{Reproducible Benchmarks:} Deterministic evaluation scripts audited with \texttt{osaudit}~\cite{monperrus2026osaudit}, tying every reported number in this paper to SHA-256 input hashes and checked-in JSON manifests.
\end{itemize}

\section{Threat Model and Verification Scope}
\label{sec:threat}

\paragraph{System Model and Mediation Boundary.}
An MCP host or orchestrator $\mathcal{H}$ runs an untrusted language-model generation loop $\mathcal{A}$ that interacts with external MCP tool servers $\{\mathcal{S}_1, \dots, \mathcal{S}_k\}$, a web/RAG retrieval pipeline $\mathcal{R}$, and a local memory database $\mathcal{M}$. To enforce complete mediation on tool traffic, $\mathcal{H}$ configures \texttt{epistemic-guard mcp-proxy} as the \texttt{stdio} launch command for every tool server $\mathcal{S}_i$, so $\mathcal{A}$ cannot communicate with $\mathcal{S}_i$ outside the proxy pipe; complete mediation holds only if $\mathcal{A}$ is granted no unproxied side channels (such as direct shell execution or built-in HTTP tools) outside those proxied servers. On turn $t$, $\mathcal{A}$ emits zero or more JSON-RPC\,2.0 \texttt{tools/call} requests mediated by \texttt{mcp-proxy}, optionally retrieves external web pages, proposes candidate memory writes to $\mathcal{M}$, and emits a candidate natural-language reply $y_t$. Before $\mathcal{H}$ displays $y_t$ or passes it downstream, $\mathcal{H}$ invokes \texttt{ReceiptSupervisor} (or CLI \texttt{verify-log}) outside $\mathcal{A}$'s prompt context.

\paragraph{Adversary and Fault Assumptions.}
We consider both stochastic model hallucinations and adversarial indirect prompt injections~\cite{greshake2023indirect} inside tool outputs or retrieved web documents:
\begin{enumerate}
  \item \emph{Post-tool action fabrication (\texttt{ASI02}, \texttt{ASI08}):} $\mathcal{A}$ may claim in $y_t$ that a side-effecting tool call succeeded when no call occurred, when the server returned \texttt{isError=true}, when argument validation rejected the call pre-execution, or when the child server crashed mid-flight. If signing keys reside inside $\mathcal{A}$'s process or child server environments, compromised code or plugins could forge receipts.
  \item \emph{Cross-session memory poisoning (\texttt{ASI06}):} $\mathcal{A}$ may attempt to persist unverified assistant narration (\texttt{LLM\_NARRATION}) or untrusted tool/web text (\texttt{TOOL\_CONTENT\_UNTRUSTED}) into $\mathcal{M}$ as verified facts, or mutate historical audit rows via SQL \texttt{UPDATE}/\texttt{DELETE} statements.
  \item \emph{Syndicated corroboration spoofing (\texttt{ASI06}, \texttt{ASI08}):} Retrieved web pages often reprint identical wire-service copy or SEO mirrors across distinct hostnames. Naive multi-URL counting treats \href{https://github.com/AlienatedAliens/epistemic-guard/blob/v0.5.2/results/table2_conformance.json\#L70}{3} mirrors of one rumor as independent corroboration for high-hazard medical, financial, or security claims.
\end{enumerate}

\paragraph{Trusted Computing Base (TCB) and Key Isolation.}
The TCB comprises the host OS kernel, the MCP host/orchestrator $\mathcal{H}$ invoking post-turn verification, the out-of-process \texttt{mcp-proxy} process holding the Ed25519 signing key, the SHA-256 digests of the \href{https://github.com/AlienatedAliens/epistemic-guard/blob/v0.5.2/results/table1_architecture.json\#L25}{16} INT8 model files (verified by default at every model load via \texttt{verify\_\allowbreak hashes=True} across explicit or OS-cache paths), and OS permissions on the SQLite store and pinned public key. When keys are injected via \texttt{EPISTEMIC\_\allowbreak GUARD\_\allowbreak MCP\_\allowbreak KEY\_\allowbreak HEX}, the secret stays in \texttt{mcp-proxy} memory and is scrubbed before spawning child servers; when \texttt{-{-}key-file} is used on disk (created atomically via \texttt{O\_\allowbreak CREAT|\allowbreak O\_\allowbreak EXCL} with POSIX mode \texttt{0600}, or restricted NTFS directory ACLs on Windows), same-host deployments against untrusted local code must run \texttt{mcp-proxy} under a dedicated OS user or container so same-UID processes cannot read \texttt{-{-}key-file}. We do \emph{not} defend against host kernel compromise, malicious MCP servers that return \texttt{ok=true} while lying about internal state, or outbound network exfiltration (where egress proxies such as Pipelock~\cite{pipelock2026} compose upstream).

\section{System Architecture}
\label{sec:arch}

\begin{table}[t]
\centering
\caption{Static architectural metrics and INT8 model footprint of \texttt{epistemic-guard} v0.5.2 (\texttt{table1\_\allowbreak architecture.json}).}
\label{tab:arch}
\footnotesize
\setlength{\tabcolsep}{4pt}
\begin{tabularx}{\columnwidth}{@{}>{\raggedright\arraybackslash}Xr@{}}
\toprule
\textbf{Architectural Property / Component} & \textbf{Value} \\
\midrule
Python source modules (\texttt{epistemic\_guard}) & \href{https://github.com/AlienatedAliens/epistemic-guard/blob/v0.5.2/results/table1_architecture.json\#L30}{15} \\
Layer contracts (\texttt{import-linter}) & \href{https://github.com/AlienatedAliens/epistemic-guard/blob/v0.5.2/results/table1_architecture.json\#L15}{1} \\
Test modules / automated test functions & \href{https://github.com/AlienatedAliens/epistemic-guard/blob/v0.5.2/results/table1_architecture.json\#L32}{4} / \href{https://github.com/AlienatedAliens/epistemic-guard/blob/v0.5.2/results/table1_architecture.json\#L33}{34} \\
MCP server tools / protocol revisions & \href{https://github.com/AlienatedAliens/epistemic-guard/blob/v0.5.2/results/table1_architecture.json\#L17}{4} / \href{https://github.com/AlienatedAliens/epistemic-guard/blob/v0.5.2/results/table1_architecture.json\#L31}{3} \\
Provenance domains / HKDF signers & \href{https://github.com/AlienatedAliens/epistemic-guard/blob/v0.5.2/results/table1_architecture.json\#L28}{7} / \href{https://github.com/AlienatedAliens/epistemic-guard/blob/v0.5.2/results/table1_architecture.json\#L14}{5} \\
Write-eligible SQLite provenance domains & \href{https://github.com/AlienatedAliens/epistemic-guard/blob/v0.5.2/results/table1_architecture.json\#L11}{5} \\
Ed25519 key / signature size (bytes) & \href{https://github.com/AlienatedAliens/epistemic-guard/blob/v0.5.2/results/table1_architecture.json\#L9}{32} / \href{https://github.com/AlienatedAliens/epistemic-guard/blob/v0.5.2/results/table1_architecture.json\#L10}{64} \\
Hazard tiers / proof standards (policy cells) & \href{https://github.com/AlienatedAliens/epistemic-guard/blob/v0.5.2/results/table1_architecture.json\#L13}{3} / \href{https://github.com/AlienatedAliens/epistemic-guard/blob/v0.5.2/results/table1_architecture.json\#L27}{4} (\href{https://github.com/AlienatedAliens/epistemic-guard/blob/v0.5.2/results/table1_architecture.json\#L26}{12}) \\
Consumer contexts / word-shingle width $k$ & \href{https://github.com/AlienatedAliens/epistemic-guard/blob/v0.5.2/results/table1_architecture.json\#L2}{4} / \href{https://github.com/AlienatedAliens/epistemic-guard/blob/v0.5.2/results/table1_architecture.json\#L29}{4} \\
\midrule
INT8 models $\times$ files (SHA-256 pinned) & \href{https://github.com/AlienatedAliens/epistemic-guard/blob/v0.5.2/results/table1_architecture.json\#L16}{4} $\times$ \href{https://github.com/AlienatedAliens/epistemic-guard/blob/v0.5.2/results/table1_architecture.json\#L12}{4} (\href{https://github.com/AlienatedAliens/epistemic-guard/blob/v0.5.2/results/table1_architecture.json\#L25}{16}) \\
\quad \texttt{action\_claim} INT8 encoder (MB) & \href{https://github.com/AlienatedAliens/epistemic-guard/blob/v0.5.2/results/table1_architecture.json\#L19}{82.67} \\
\quad \texttt{hazard} INT8 classifier (MB) & \href{https://github.com/AlienatedAliens/epistemic-guard/blob/v0.5.2/results/table1_architecture.json\#L20}{82.65} \\
\quad \texttt{nli} DeBERTa-v3 cross-encoder (MB) & \href{https://github.com/AlienatedAliens/epistemic-guard/blob/v0.5.2/results/table1_architecture.json\#L21}{91.1} \\
\quad \texttt{web\_evidence} MiniLM encoder (MB) & \href{https://github.com/AlienatedAliens/epistemic-guard/blob/v0.5.2/results/table1_architecture.json\#L22}{23.06} \\
\textbf{Total unpacked INT8 bundle (MB)} & \textbf{\href{https://github.com/AlienatedAliens/epistemic-guard/blob/v0.5.2/results/table1_architecture.json\#L34}{279.48}} \\
\bottomrule
\end{tabularx}
\end{table}

Table~\ref{tab:arch} lists the module counts and model sizes of \texttt{epistemic-guard}. The package contains \href{https://github.com/AlienatedAliens/epistemic-guard/blob/v0.5.2/results/table1_architecture.json\#L30}{15} typed Python modules (\texttt{mypy -{-}strict}) checked by \href{https://github.com/AlienatedAliens/epistemic-guard/blob/v0.5.2/results/table1_architecture.json\#L15}{1} \texttt{import-linter} layer contract so that cryptographic verification, supervision, and storage modules never import from CLI, MCP server, or integration layers.

\subsection{Domain-Separated Cryptographic Provenance and AER-1 Receipts}
\label{sec:crypto}

Rather than sharing one symmetric key across components, \texttt{provenance.py} derives \href{https://github.com/AlienatedAliens/epistemic-guard/blob/v0.5.2/results/table1_architecture.json\#L14}{5} domain-isolated RFC\,8032 Ed25519~\cite{josefsson2017rfc8032} keypairs (\href{https://github.com/AlienatedAliens/epistemic-guard/blob/v0.5.2/results/table1_architecture.json\#L9}{32}-byte keys, \href{https://github.com/AlienatedAliens/epistemic-guard/blob/v0.5.2/results/table1_architecture.json\#L10}{64}-byte signatures) from a master seed via RFC\,5869 HKDF-SHA256~\cite{krawczyk2010rfc5869} across \href{https://github.com/AlienatedAliens/epistemic-guard/blob/v0.5.2/results/table1_architecture.json\#L28}{7} provenance domains (\texttt{Provenance}): \texttt{USER\_\allowbreak EXPLICIT}, \texttt{USER\_\allowbreak CONFIRMED}, \texttt{TOOL\_\allowbreak VERIFIED}, \texttt{LISP\_\allowbreak ASSERTION}, \texttt{INFERRED}, \texttt{LLM\_\allowbreak NARRATION}, and \texttt{TOOL\_\allowbreak CONTENT\_\allowbreak UNTRUSTED}. \texttt{Signer\allowbreak Factory} provides no signing constructor for \texttt{LLM\_\allowbreak NARRATION} or \texttt{TOOL\_\allowbreak CONTENT\_\allowbreak UNTRUSTED}, so model prose and untrusted tool text cannot produce a signed envelope through the public API.

Each tool call produces a canonical \texttt{Tool\allowbreak Receipt} (serialized with sorted keys and \texttt{allow\_nan=False} so \texttt{NaN}/\texttt{Infinity} floats cannot break cross-runtime canonicalization) containing the call ID, tool name, structured arguments, SHA-256 output digest, execution flag (\texttt{executed}), outcome flag (\texttt{ok}), error code, optional session ID, UTC timestamp, and detached Ed25519 signature. Receipts append into an RFC\,6962~\cite{laurie2013rfc6962} domain-separated Merkle tree (\texttt{0x00} leaf prefix, \texttt{0x01} interior node prefix) with logarithmic inclusion proofs and path-side geometry checks. For cross-tool interoperability, \texttt{Tool\allowbreak Receipt.\allowbreak to\_\allowbreak aer1} and \texttt{Tool\allowbreak Receipt.\allowbreak from\_\allowbreak aer1} implement single-receipt translation and core/binding verification against \texttt{AER-1} (\texttt{draft-zambo-aer1-26}, Sections\,3--6, \texttt{receipt\_\allowbreak schema\_\allowbreak version} \texttt{0.4} with \texttt{0.2}/\texttt{0.3} legacy read support)~\cite{zambo2026aer1}, committing execution state, input arguments, call/session IDs, and SHA-256 input/output digests while multi-entry sequencing and detached Ed25519 signatures remain in \texttt{Receipt\allowbreak Log\allowbreak Entry} and \texttt{Merkle\allowbreak Receipt\allowbreak Log}.

\subsection{Out-of-Process \texttt{stdio} MCP Receipt Proxy}
\label{sec:proxy}

In-process Python decorators (\texttt{Receipt\allowbreak Recorder}) protect benign agent loops against hallucinated actions, but share memory and environment variables with the agent runtime. For untrusted or prompt-injected runtimes, \texttt{epistemic-guard mcp-proxy} interposes as an out-of-process \texttt{stdio} parent between the MCP host and any downstream MCP tool server:
\begin{enumerate}
  \item \textbf{Secret Scrubbing:} Before spawning the child MCP server, \texttt{mcp-proxy} case-insensitively strips signing-secret environment variables (including \texttt{EPISTEMIC\_\allowbreak GUARD\_\allowbreak MASTER\_\allowbreak KEY} and \texttt{EPISTEMIC\_\allowbreak GUARD\_\allowbreak MCP\_\allowbreak KEY\_\allowbreak HEX}) from the child's environment block so compromised child servers cannot forge receipts.
  \item \textbf{Four-State Wire Classification:} As JSON-RPC\,2.0 frames stream over \texttt{stdin}/\texttt{stdout}, \texttt{mcp-proxy} correlates \texttt{tools/call} requests in FIFO order per JSON-RPC \texttt{id} and distinguishes \href{https://github.com/AlienatedAliens/epistemic-guard/blob/v0.5.2/results/table2_conformance.json\#L31}{4} runtime outcomes: (a) normal tool completion (\texttt{executed=true, ok=true}); (b) application-level MCP tool failure where \texttt{result.isError=true} (\texttt{executed=true, ok=false, mcp\_\allowbreak tool\_\allowbreak error}); (c) pre-execution rejection on standard JSON-RPC protocol/parameter error codes (\texttt{-32700}, \texttt{-32600}, \texttt{-32601}, \texttt{-32602}) or schema-validation error markers (\texttt{executed=false, ok=false}); and (d) child server crash or EOF mid-call (\texttt{executed=true, ok=false, mcp\_\allowbreak server\_\allowbreak eof}). Non-standard server errors default conservatively to \texttt{executed=true, ok=false} so neither rejected nor failed calls can ground completion prose.
  \item \textbf{Key-Pinned Append-Only Audit Log:} Each \texttt{Receipt\allowbreak Log\allowbreak Entry} carries both an inner Ed25519 \texttt{Tool\allowbreak Receipt} signature and a domain-separated Ed25519 \texttt{entry\_\allowbreak signature} over \texttt{(seq, session\_\allowbreak id, leaf\_\allowbreak sha256, prev\_\allowbreak chain\_\allowbreak sha256, chain\_\allowbreak sha256, merkle\_\allowbreak root)}, acquires cross-process OS file locks for multi-worker concurrency, chains optionally across rotated Ed25519 keys (\texttt{-{-}prev-log} / \texttt{-{-}prev-pubkey-file}), and flushes immediately to an append-only JSONL Merkle log; \texttt{verify-log} (with external \texttt{-{-}pubkey-file} pinning) rejects payload or embedded-\texttt{AER-1} mutation, active reorder-and-rehash attacks, duplicate call IDs, cross-session replay, and key substitution, while \texttt{attach\_\allowbreak otel\_\allowbreak evidence} exports receipt digests to OpenTelemetry traces.
\end{enumerate}
In addition to \texttt{mcp-proxy}, \texttt{epistemic-guard mcp-server} exposes \href{https://github.com/AlienatedAliens/epistemic-guard/blob/v0.5.2/results/table1_architecture.json\#L17}{4} verification tools directly over MCP across \href{https://github.com/AlienatedAliens/epistemic-guard/blob/v0.5.2/results/table1_architecture.json\#L31}{3} protocol revisions (\texttt{2024-11-05}, \texttt{2025-03-26}, and \texttt{2025-06-18}).

\subsection{Hybrid Symbolic--Neural Post-Turn Prose Supervision}
\label{sec:supervision}

Regex-only action supervisors miss indirect phrasing, nominalizations, and passive-voice state mutations, whereas off-the-shelf NLI encoders treat unexecuted tool claims against empty receipt logs as neutral~\cite{maxwell2026evasion}. Across the full package, \texttt{epistemic-guard} distributes \href{https://github.com/AlienatedAliens/epistemic-guard/blob/v0.5.2/results/table1_architecture.json\#L16}{4} OpenVINO INT8 models (\href{https://github.com/AlienatedAliens/epistemic-guard/blob/v0.5.2/results/table1_architecture.json\#L25}{16} SHA-256-pinned files totaling \href{https://github.com/AlienatedAliens/epistemic-guard/blob/v0.5.2/results/table1_architecture.json\#L34}{279.48}\,MB over HTTPS from Hugging Face Hub or GitHub Releases, extracted with symlink/path-traversal guards and verified by SHA-256 at every load). Only \texttt{action\_claim} (\href{https://github.com/AlienatedAliens/epistemic-guard/blob/v0.5.2/results/table1_architecture.json\#L19}{82.67}\,MB) runs on the post-turn supervision path, where \texttt{ReceiptSupervisor} executes a two-stage CPU pipeline:
\begin{enumerate}
  \item \textbf{Speech-Act Action-Claim Detection:} Non-assertive spans (negations, conditionals, attributed quotes, creative/translation prompts, and fenced code blocks) are masked while optional Structured Output Action Contract (\texttt{SOAC}, explicit \texttt{[ACTION: verb target]} tags~\cite{maxwell2026evasion}) declarations are extracted, and candidate clauses are scored by the INT8 \texttt{action\_claim} cross-encoder. If no completed external side effect is asserted, the turn passes without requiring receipts.
  \item \textbf{Cryptographic Receipt \& Target Matching:} When a side-effect claim is detected, \texttt{Receipt\allowbreak Supervisor} verifies the turn's \texttt{Tool\allowbreak Receipt} set via \texttt{Provenance\allowbreak Gate} (enforcing Ed25519 signatures, session binding, and receipt-age bounds when configured), excludes any receipt with \texttt{executed=false} or \texttt{ok=false}, and matches claimed action families, additive claim counts, and concrete file/URL/email/identifier targets against verified receipt \texttt{args} on path and token boundaries (rejecting substring collisions and bare zero-argument receipts when prose names specific targets). Ungrounded claims return \texttt{Supervision\allowbreak Verdict(\allowbreak is\_valid=false)} with character-offset spans and \texttt{to\_\allowbreak retry\_\allowbreak prompt()} remediation text.
\end{enumerate}

\subsection{Provenance-Gated SQLite Memory and De-Syndicated RAG}
\label{sec:memory}

\paragraph{Storage-Boundary Write Gate.}
In \texttt{store.py}, \texttt{SQLite\allowbreak Provenance\allowbreak Store} wraps SQLite WAL persistence behind mandatory \texttt{ProvenanceGate} verification and serialized \texttt{BEGIN IMMEDIATE} write transactions. Only \href{https://github.com/AlienatedAliens/epistemic-guard/blob/v0.5.2/results/table1_architecture.json\#L11}{5} provenance domains with valid detached Ed25519 signatures are write-eligible; writes carrying \texttt{LLM\_NARRATION}, \texttt{TOOL\_\allowbreak CONTENT\_\allowbreak UNTRUSTED}, forged keys, or mutated payloads raise \texttt{ProvenanceError} or \texttt{Signature\allowbreak Tamper\allowbreak Error} before touching SQLite. Accepted envelopes link into a cumulative SHA-256 chain digest protected by \href{https://github.com/AlienatedAliens/epistemic-guard/blob/v0.5.2/results/table2_conformance.json\#L53}{2} SQL \texttt{BEFORE UPDATE} and \texttt{BEFORE DELETE} triggers that abort in-place mutation of committed rows.

\paragraph{Syndication Deduplication and Hazard-Tiered Corroboration.}
Before web-retrieved claims are promoted into \texttt{INFERRED} memory or stated without hedging, \texttt{corroboration.py} shards candidate passages into word shingles of width $k=\href{https://github.com/AlienatedAliens/epistemic-guard/blob/v0.5.2/results/table1_architecture.json\#L29}{4}$, drops non-evidentiary boilerplate with the MiniLM~\cite{wang2020minilm} INT8 \texttt{web\_evidence} encoder (\href{https://github.com/AlienatedAliens/epistemic-guard/blob/v0.5.2/results/table1_architecture.json\#L22}{23.06}\,MB), checks cross-source contradictions with the DeBERTa-v3~\cite{he2023debertav3} INT8 \texttt{nli} cross-encoder (\href{https://github.com/AlienatedAliens/epistemic-guard/blob/v0.5.2/results/table1_architecture.json\#L21}{91.1}\,MB), and merges near-duplicate mirrors across registrable domains using explicit similarity gates: anchor-token Jaccard $\ge \href{https://github.com/AlienatedAliens/epistemic-guard/blob/v0.5.2/results/table1_architecture.json\#L7}{0.45}$, $3$-shingle Jaccard $\ge \href{https://github.com/AlienatedAliens/epistemic-guard/blob/v0.5.2/results/table1_architecture.json\#L4}{0.25}$, $4$-shingle Jaccard $\ge \href{https://github.com/AlienatedAliens/epistemic-guard/blob/v0.5.2/results/table1_architecture.json\#L5}{0.55}$~\cite{broder1997resemblance}, or 64-bit SimHash Hamming distance $\le \href{https://github.com/AlienatedAliens/epistemic-guard/blob/v0.5.2/results/table1_architecture.json\#L6}{12}$~\cite{charikar2002similarity}. Independent source counts are checked against \href{https://github.com/AlienatedAliens/epistemic-guard/blob/v0.5.2/results/table1_architecture.json\#L2}{4} consumer action contexts (\texttt{casual\_\allowbreak mention}, \texttt{memory\_\allowbreak commit\_\allowbreak inferred}, \texttt{unhedged\_\allowbreak assertion}, and \texttt{irreversible\_\allowbreak action}) and a \href{https://github.com/AlienatedAliens/epistemic-guard/blob/v0.5.2/results/table1_architecture.json\#L13}{3}-tier hazard classifier (\href{https://github.com/AlienatedAliens/epistemic-guard/blob/v0.5.2/results/table1_architecture.json\#L20}{82.65}\,MB; \texttt{LOW}, \texttt{MEDIUM}, \texttt{HIGH}) mapped onto \href{https://github.com/AlienatedAliens/epistemic-guard/blob/v0.5.2/results/table1_architecture.json\#L27}{4} proof standards (\href{https://github.com/AlienatedAliens/epistemic-guard/blob/v0.5.2/results/table1_architecture.json\#L26}{12} policy cells). High-hazard claims backed only by wire-syndicated mirrors of one origin collapse to one cluster and fail corroboration.

\section{Empirical Evaluation}
\label{sec:eval}

We structure our evaluation around two research questions:
\begin{itemize}
  \item \textbf{RQ1 (Deterministic Conformance, Concurrency/Replay Faults, and Paraphrase Sensitivity):} Do the Merkle receipt ledger, \texttt{AER-1} interop layer, out-of-process \texttt{mcp-proxy} wire classifier, concurrent log writer, and SQLite provenance gate enforce their deterministic invariants under fault injection, and at what rewrite intensity does shingle/SimHash syndication deduplication degrade?
  \item \textbf{RQ2 (Prose Supervision Accuracy, Live MCP Receipt Binding, \& Resource Cost):} How accurately and rapidly does \texttt{ReceiptSupervisor} detect ungrounded post-turn action claims across frozen benchmark cohorts ($N=\href{https://github.com/AlienatedAliens/epistemic-guard/blob/v0.5.2/results/table3_heldout_supervision.json\#L175}{250}$) and live end-to-end \texttt{mcp-proxy} fault-injection traces ($N=\href{https://github.com/AlienatedAliens/epistemic-guard/blob/v0.5.2/results/table3_heldout_supervision.json\#L136}{60}$) compared with stage ablations, 12B--14B GPU LLM judges, and off-the-shelf CPU NLI baselines?
\end{itemize}

All experiments are deterministic, version-pinned, and audited with \texttt{osaudit}~\cite{monperrus2026osaudit} manifests under \texttt{results/} (\texttt{table1\_\allowbreak architecture.json}, \texttt{table2\_\allowbreak conformance.json}, and \texttt{table3\_\allowbreak heldout\_\allowbreak supervision.json}). Binomial proportions use two-sided \href{https://github.com/AlienatedAliens/epistemic-guard/blob/v0.5.2/osaudit.toml}{95\%} Wilson score confidence intervals~\cite{wilson1927probable}, and paired classifier comparisons use exact two-sided McNemar tests~\cite{mcnemar1947note}.

\subsection{RQ1: Conformance, Fault-Injection, and Paraphrase Degradation}
\label{sec:rq1}

\begin{table*}[t]
\centering
\caption{Deterministic protocol conformance, multi-worker concurrency, and tamper/replay-rejection results across runtime enforcement subsystems (\texttt{run\_protocol\_conformance.py} $\rightarrow$ \texttt{results/table2\_conformance.json}).}
\label{tab:conformance}
\footnotesize
\begin{tabularx}{\textwidth}{@{}l>{\raggedright\arraybackslash}Xrrr@{}}
\toprule
\textbf{Subsystem} & \textbf{Evaluated Invariant / Fault-Injection Workload} & \textbf{Trials} & \textbf{Verified / Blocked} & \textbf{Rate} \\
\midrule
RFC\,6962 Merkle Ledger & Valid leaf inclusion proofs verified & \href{https://github.com/AlienatedAliens/epistemic-guard/blob/v0.5.2/results/table2_conformance.json\#L43}{200} & \href{https://github.com/AlienatedAliens/epistemic-guard/blob/v0.5.2/results/table2_conformance.json\#L47}{200} & \href{https://github.com/AlienatedAliens/epistemic-guard/blob/v0.5.2/results/table2_conformance.json\#L46}{100.0\%} \\
RFC\,6962 Merkle Ledger & Mutated receipt payloads rejected (\href{https://github.com/AlienatedAliens/epistemic-guard/blob/v0.5.2/results/table2_conformance.json\#L42}{0} false accepts) & \href{https://github.com/AlienatedAliens/epistemic-guard/blob/v0.5.2/results/table2_conformance.json\#L43}{200} & \href{https://github.com/AlienatedAliens/epistemic-guard/blob/v0.5.2/results/table2_conformance.json\#L45}{200} & \href{https://github.com/AlienatedAliens/epistemic-guard/blob/v0.5.2/results/table2_conformance.json\#L44}{100.0\%} \\
\texttt{AER-1} Interoperability & Official \texttt{draft-zambo-aer1-26} ingestion kit (\href{https://github.com/AlienatedAliens/epistemic-guard/blob/v0.5.2/results/table2_conformance.json\#L19}{22} valid, \href{https://github.com/AlienatedAliens/epistemic-guard/blob/v0.5.2/results/table2_conformance.json\#L16}{36} invalid) & \href{https://github.com/AlienatedAliens/epistemic-guard/blob/v0.5.2/results/table2_conformance.json\#L21}{58} & \href{https://github.com/AlienatedAliens/epistemic-guard/blob/v0.5.2/results/table2_conformance.json\#L18}{58} & \href{https://github.com/AlienatedAliens/epistemic-guard/blob/v0.5.2/results/table2_conformance.json\#L15}{100.0\%} \\
\texttt{AER-1} Interoperability & Reference-verified \texttt{to\_aer1()} exports, round-trip, \& tampered-digest rejection & \href{https://github.com/AlienatedAliens/epistemic-guard/blob/v0.5.2/results/table2_conformance.json\#L14}{50} & \href{https://github.com/AlienatedAliens/epistemic-guard/blob/v0.5.2/results/table2_conformance.json\#L22}{50} & \href{https://github.com/AlienatedAliens/epistemic-guard/blob/v0.5.2/results/table2_conformance.json\#L23}{100.0\%} \\
Out-of-Process \texttt{mcp-proxy} & Wire calls classified across \href{https://github.com/AlienatedAliens/epistemic-guard/blob/v0.5.2/results/table2_conformance.json\#L31}{4} modes (\href{https://github.com/AlienatedAliens/epistemic-guard/blob/v0.5.2/results/table2_conformance.json\#L33}{10}/mode; \href{https://github.com/AlienatedAliens/epistemic-guard/blob/v0.5.2/results/table2_conformance.json\#L28}{0} env key leaks) & \href{https://github.com/AlienatedAliens/epistemic-guard/blob/v0.5.2/results/table2_conformance.json\#L34}{40} & \href{https://github.com/AlienatedAliens/epistemic-guard/blob/v0.5.2/results/table2_conformance.json\#L38}{40} & \href{https://github.com/AlienatedAliens/epistemic-guard/blob/v0.5.2/results/table2_conformance.json\#L39}{100.0\%} \\
Concurrent Audit Log & Multi-worker locked appends verified (\href{https://github.com/AlienatedAliens/epistemic-guard/blob/v0.5.2/results/table2_conformance.json\#L6}{4} threads $\times$ 25 appends) & \href{https://github.com/AlienatedAliens/epistemic-guard/blob/v0.5.2/results/table2_conformance.json\#L5}{100} & \href{https://github.com/AlienatedAliens/epistemic-guard/blob/v0.5.2/results/table2_conformance.json\#L3}{100} & \href{https://github.com/AlienatedAliens/epistemic-guard/blob/v0.5.2/results/table2_conformance.json\#L4}{100.0\%} \\
Replay \& Canonicalization & Cross-session replay, entry reordering, and \texttt{NaN}/\texttt{Inf} floats rejected & \href{https://github.com/AlienatedAliens/epistemic-guard/blob/v0.5.2/results/table2_conformance.json\#L9}{4} & \href{https://github.com/AlienatedAliens/epistemic-guard/blob/v0.5.2/results/table2_conformance.json\#L8}{4} & \href{https://github.com/AlienatedAliens/epistemic-guard/blob/v0.5.2/results/table2_conformance.json\#L7}{100.0\%} \\
SQLite Provenance Gate & Authorized signed domain writes accepted (\href{https://github.com/AlienatedAliens/epistemic-guard/blob/v0.5.2/results/table2_conformance.json\#L51}{0} false rejections) & \href{https://github.com/AlienatedAliens/epistemic-guard/blob/v0.5.2/results/table2_conformance.json\#L59}{200} & \href{https://github.com/AlienatedAliens/epistemic-guard/blob/v0.5.2/results/table2_conformance.json\#L63}{200} & \href{https://github.com/AlienatedAliens/epistemic-guard/blob/v0.5.2/results/table2_conformance.json\#L62}{100.0\%} \\
SQLite Provenance Gate & Unauthorized writes blocked (\href{https://github.com/AlienatedAliens/epistemic-guard/blob/v0.5.2/results/table2_conformance.json\#L56}{50} \texttt{LLM\_NARRATION} + \href{https://github.com/AlienatedAliens/epistemic-guard/blob/v0.5.2/results/table2_conformance.json\#L55}{50} forged keys) & \href{https://github.com/AlienatedAliens/epistemic-guard/blob/v0.5.2/results/table2_conformance.json\#L58}{100} & \href{https://github.com/AlienatedAliens/epistemic-guard/blob/v0.5.2/results/table2_conformance.json\#L61}{100} & \href{https://github.com/AlienatedAliens/epistemic-guard/blob/v0.5.2/results/table2_conformance.json\#L60}{100.0\%} \\
SQLite Immutability & Direct SQL \texttt{UPDATE} and \texttt{DELETE} mutations aborted by triggers & \href{https://github.com/AlienatedAliens/epistemic-guard/blob/v0.5.2/results/table2_conformance.json\#L53}{2} & \href{https://github.com/AlienatedAliens/epistemic-guard/blob/v0.5.2/results/table2_conformance.json\#L52}{2} & \href{https://github.com/AlienatedAliens/epistemic-guard/blob/v0.5.2/results/table2_conformance.json\#L54}{100.0\%} \\
RAG Syndication Filter & \href{https://github.com/AlienatedAliens/epistemic-guard/blob/v0.5.2/results/table2_conformance.json\#L81}{135} wire-mirror URLs (\href{https://github.com/AlienatedAliens/epistemic-guard/blob/v0.5.2/results/table2_conformance.json\#L70}{3}/cluster) collapsed; high-hazard promotion blocked & \href{https://github.com/AlienatedAliens/epistemic-guard/blob/v0.5.2/results/table2_conformance.json\#L71}{45} & \href{https://github.com/AlienatedAliens/epistemic-guard/blob/v0.5.2/results/table2_conformance.json\#L69}{45} & \href{https://github.com/AlienatedAliens/epistemic-guard/blob/v0.5.2/results/table2_conformance.json\#L68}{100.0\%} \\
\bottomrule
\end{tabularx}
\end{table*}

Table~\ref{tab:conformance} reports deterministic conformance and fault-injection results from \texttt{run\_\allowbreak protocol\_\allowbreak conformance.py}:
\begin{itemize}
  \item \textbf{Merkle, \texttt{AER-1}, \& Replay Integrity:} Across \href{https://github.com/AlienatedAliens/epistemic-guard/blob/v0.5.2/results/table2_conformance.json\#L43}{200} signed receipts, all \href{https://github.com/AlienatedAliens/epistemic-guard/blob/v0.5.2/results/table2_conformance.json\#L47}{200} RFC\,6962 inclusion proofs verify (\href{https://github.com/AlienatedAliens/epistemic-guard/blob/v0.5.2/results/table2_conformance.json\#L46}{100.0\%}) and all \href{https://github.com/AlienatedAliens/epistemic-guard/blob/v0.5.2/results/table2_conformance.json\#L45}{200} payload mutations raise \texttt{MerkleTamperError} (\href{https://github.com/AlienatedAliens/epistemic-guard/blob/v0.5.2/results/table2_conformance.json\#L42}{0} false accepts). Evaluated against the official \texttt{draft-zambo-aer1-26} conformance kit~\cite{zambo2026aer1}, \texttt{ToolReceipt.from\_aer1} matches \href{https://github.com/AlienatedAliens/epistemic-guard/blob/v0.5.2/results/table2_conformance.json\#L18}{58}/\href{https://github.com/AlienatedAliens/epistemic-guard/blob/v0.5.2/results/table2_conformance.json\#L21}{58} single-receipt ingestion vectors (\href{https://github.com/AlienatedAliens/epistemic-guard/blob/v0.5.2/results/table2_conformance.json\#L15}{100.0\%}: \href{https://github.com/AlienatedAliens/epistemic-guard/blob/v0.5.2/results/table2_conformance.json\#L19}{22}/\href{https://github.com/AlienatedAliens/epistemic-guard/blob/v0.5.2/results/table2_conformance.json\#L20}{22} valid/profile-boundary accepted, \href{https://github.com/AlienatedAliens/epistemic-guard/blob/v0.5.2/results/table2_conformance.json\#L16}{36}/\href{https://github.com/AlienatedAliens/epistemic-guard/blob/v0.5.2/results/table2_conformance.json\#L17}{36} invalid rejected), while all \href{https://github.com/AlienatedAliens/epistemic-guard/blob/v0.5.2/results/table2_conformance.json\#L22}{50}/\href{https://github.com/AlienatedAliens/epistemic-guard/blob/v0.5.2/results/table2_conformance.json\#L14}{50} \texttt{ToolReceipt.to\_aer1} exports pass the kit's reference verifier (\href{https://github.com/AlienatedAliens/epistemic-guard/blob/v0.5.2/results/table2_conformance.json\#L23}{100.0\%} across core, version-binding, and reference-producer checks), round-trip cleanly (\href{https://github.com/AlienatedAliens/epistemic-guard/blob/v0.5.2/results/table2_conformance.json\#L25}{50}/\href{https://github.com/AlienatedAliens/epistemic-guard/blob/v0.5.2/results/table2_conformance.json\#L14}{50}, \href{https://github.com/AlienatedAliens/epistemic-guard/blob/v0.5.2/results/table2_conformance.json\#L24}{100.0\%}), and reject digest edits (\href{https://github.com/AlienatedAliens/epistemic-guard/blob/v0.5.2/results/table2_conformance.json\#L12}{50}/\href{https://github.com/AlienatedAliens/epistemic-guard/blob/v0.5.2/results/table2_conformance.json\#L14}{50}, \href{https://github.com/AlienatedAliens/epistemic-guard/blob/v0.5.2/results/table2_conformance.json\#L13}{100.0\%}). Under \href{https://github.com/AlienatedAliens/epistemic-guard/blob/v0.5.2/results/table2_conformance.json\#L6}{4}-thread contention (\href{https://github.com/AlienatedAliens/epistemic-guard/blob/v0.5.2/results/table2_conformance.json\#L5}{100} concurrent appends to one log), OS file locking preserves sequential chain hashes across all \href{https://github.com/AlienatedAliens/epistemic-guard/blob/v0.5.2/results/table2_conformance.json\#L3}{100} entries (\href{https://github.com/AlienatedAliens/epistemic-guard/blob/v0.5.2/results/table2_conformance.json\#L4}{100.0\%}), while \href{https://github.com/AlienatedAliens/epistemic-guard/blob/v0.5.2/results/table2_conformance.json\#L8}{4}/\href{https://github.com/AlienatedAliens/epistemic-guard/blob/v0.5.2/results/table2_conformance.json\#L9}{4} cross-session replay, active rehashed-reorder, and \texttt{NaN}/\texttt{Infinity} canonicalization attempts fail closed (\href{https://github.com/AlienatedAliens/epistemic-guard/blob/v0.5.2/results/table2_conformance.json\#L7}{100.0\%}).
  \item \textbf{Out-of-Process MCP Wire Fault Injection:} Spawning live child \texttt{stdio} MCP servers under \texttt{mcp-proxy} across \href{https://github.com/AlienatedAliens/epistemic-guard/blob/v0.5.2/results/table2_conformance.json\#L34}{40} tool invocations (\href{https://github.com/AlienatedAliens/epistemic-guard/blob/v0.5.2/results/table2_conformance.json\#L33}{10} each across normal completion, \texttt{result.isError=true}, JSON-RPC \texttt{-32602} invalid-parameter rejection, and mid-request child crash), all \href{https://github.com/AlienatedAliens/epistemic-guard/blob/v0.5.2/results/table2_conformance.json\#L38}{40} logged receipts pass chain and pinned-key verification (\href{https://github.com/AlienatedAliens/epistemic-guard/blob/v0.5.2/results/table2_conformance.json\#L32}{1} pinned key, \href{https://github.com/AlienatedAliens/epistemic-guard/blob/v0.5.2/results/table2_conformance.json\#L36}{1}/\href{https://github.com/AlienatedAliens/epistemic-guard/blob/v0.5.2/results/table2_conformance.json\#L37}{1} post-hoc log edit rejected), separating \href{https://github.com/AlienatedAliens/epistemic-guard/blob/v0.5.2/results/table2_conformance.json\#L30}{30} dispatched/crash calls (\href{https://github.com/AlienatedAliens/epistemic-guard/blob/v0.5.2/results/table2_conformance.json\#L35}{10} \texttt{ok=true}) from \href{https://github.com/AlienatedAliens/epistemic-guard/blob/v0.5.2/results/table2_conformance.json\#L29}{10} pre-execution rejections (\texttt{executed=false}) with \href{https://github.com/AlienatedAliens/epistemic-guard/blob/v0.5.2/results/table2_conformance.json\#L28}{0} child-environment key leaks.
  \item \textbf{SQLite Write Gate \& Paraphrase Sensitivity:} Across \href{https://github.com/AlienatedAliens/epistemic-guard/blob/v0.5.2/results/table2_conformance.json\#L57}{300} storage writes, all \href{https://github.com/AlienatedAliens/epistemic-guard/blob/v0.5.2/results/table2_conformance.json\#L63}{200} authorized envelopes verify along the \href{https://github.com/AlienatedAliens/epistemic-guard/blob/v0.5.2/results/table2_conformance.json\#L64}{200}-row hash chain (\href{https://github.com/AlienatedAliens/epistemic-guard/blob/v0.5.2/results/table2_conformance.json\#L50}{0.0\%} false rejections), all \href{https://github.com/AlienatedAliens/epistemic-guard/blob/v0.5.2/results/table2_conformance.json\#L61}{100} unauthorized writes (\href{https://github.com/AlienatedAliens/epistemic-guard/blob/v0.5.2/results/table2_conformance.json\#L56}{50} \texttt{LLM\_NARRATION} + \href{https://github.com/AlienatedAliens/epistemic-guard/blob/v0.5.2/results/table2_conformance.json\#L55}{50} forged keys) are blocked (\href{https://github.com/AlienatedAliens/epistemic-guard/blob/v0.5.2/results/table2_conformance.json\#L60}{100.0\%}), and \href{https://github.com/AlienatedAliens/epistemic-guard/blob/v0.5.2/results/table2_conformance.json\#L52}{2}/\href{https://github.com/AlienatedAliens/epistemic-guard/blob/v0.5.2/results/table2_conformance.json\#L53}{2} SQL mutations trigger \texttt{IntegrityError}. Across \href{https://github.com/AlienatedAliens/epistemic-guard/blob/v0.5.2/results/table2_conformance.json\#L71}{45} wire-syndication clusters (\href{https://github.com/AlienatedAliens/epistemic-guard/blob/v0.5.2/results/table2_conformance.json\#L81}{135} URLs, \href{https://github.com/AlienatedAliens/epistemic-guard/blob/v0.5.2/results/table2_conformance.json\#L70}{3} domains/cluster), all \href{https://github.com/AlienatedAliens/epistemic-guard/blob/v0.5.2/results/table2_conformance.json\#L80}{90} duplicate mirrors (\href{https://github.com/AlienatedAliens/epistemic-guard/blob/v0.5.2/results/table2_conformance.json\#L67}{2}/cluster) collapse to single-origin clusters (\href{https://github.com/AlienatedAliens/epistemic-guard/blob/v0.5.2/results/table2_conformance.json\#L73}{45}/\href{https://github.com/AlienatedAliens/epistemic-guard/blob/v0.5.2/results/table2_conformance.json\#L71}{45}, \href{https://github.com/AlienatedAliens/epistemic-guard/blob/v0.5.2/results/table2_conformance.json\#L74}{100.0\%}). To locate the lexical evasion boundary, a 5-tier sensitivity sweep of \href{https://github.com/AlienatedAliens/epistemic-guard/blob/v0.5.2/results/table2_conformance.json\#L72}{50} clusters per rewrite tier shows single-origin collapse holds at \href{https://github.com/AlienatedAliens/epistemic-guard/blob/v0.5.2/results/table2_conformance.json\#L75}{100.0\%} under dateline/footer wrappers and \href{https://github.com/AlienatedAliens/epistemic-guard/blob/v0.5.2/results/table2_conformance.json\#L77}{100.0\%} under light ($\sim$15\%) lexical edits, tapers to \href{https://github.com/AlienatedAliens/epistemic-guard/blob/v0.5.2/results/table2_conformance.json\#L76}{40.0\%} on intermediate ($\sim$30\%) lexical edits, and drops to \href{https://github.com/AlienatedAliens/epistemic-guard/blob/v0.5.2/results/table2_conformance.json\#L78}{0.0\%} on moderate ($\sim$45\%) clause restructuring and \href{https://github.com/AlienatedAliens/epistemic-guard/blob/v0.5.2/results/table2_conformance.json\#L79}{0.0\%} on heavy ($>$80\%) abstractive rewrites, confirming that $k=\href{https://github.com/AlienatedAliens/epistemic-guard/blob/v0.5.2/results/table1_architecture.json\#L29}{4}$ shingling targets verbatim and lightly edited wire syndication rather than semantic paraphrase.
\end{itemize}

\subsection{RQ2: Frozen-Benchmark and Live MCP Supervision}
\label{sec:rq2}

\begin{table*}[t]
\centering
\caption{Post-turn action-claim supervision on the $N=\href{https://github.com/AlienatedAliens/epistemic-guard/blob/v0.5.2/results/table3_heldout_supervision.json\#L175}{250}$ frozen benchmark (\href{https://github.com/AlienatedAliens/epistemic-guard/blob/v0.5.2/results/table3_heldout_supervision.json\#L174}{122} ungrounded, \href{https://github.com/AlienatedAliens/epistemic-guard/blob/v0.5.2/results/table3_heldout_supervision.json\#L173}{128} grounded/benign; tool-less \texttt{Org-100} sub-cohort) and $N=\href{https://github.com/AlienatedAliens/epistemic-guard/blob/v0.5.2/results/table3_heldout_supervision.json\#L136}{60}$ live \texttt{mcp-proxy} traces (\texttt{Live-60}; latencies in ms, GPU VRAM in MiB; \texttt{results/table3\_heldout\_supervision.json}).}
\label{tab:heldout}
\footnotesize
\setlength{\tabcolsep}{0pt}
\begin{tabular*}{\textwidth}{@{\extracolsep{\fill}}lrrrrrrrr@{}}
\toprule
\textbf{Supervisor / Baseline} & \textbf{Accuracy (\href{https://github.com/AlienatedAliens/epistemic-guard/blob/v0.5.2/osaudit.toml}{95\%} CI)} & \textbf{Recall} & \textbf{FPR} & \textbf{Org-100} & \textbf{Live-60} & \textbf{p50} & \textbf{p95} & \textbf{VRAM} \\
\midrule
\textbf{\texttt{epistemic-guard} INT8 (CPU)} & \textbf{\href{https://github.com/AlienatedAliens/epistemic-guard/blob/v0.5.2/results/table3_heldout_supervision.json\#L71}{93.2\%}} [\href{https://github.com/AlienatedAliens/epistemic-guard/blob/v0.5.2/results/table3_heldout_supervision.json\#L77}{89.38\%}, \href{https://github.com/AlienatedAliens/epistemic-guard/blob/v0.5.2/results/table3_heldout_supervision.json\#L76}{95.71\%}] & \href{https://github.com/AlienatedAliens/epistemic-guard/blob/v0.5.2/results/table3_heldout_supervision.json\#L96}{96.72\%} & \textbf{\href{https://github.com/AlienatedAliens/epistemic-guard/blob/v0.5.2/results/table3_heldout_supervision.json\#L83}{10.16\%}} & \textbf{\href{https://github.com/AlienatedAliens/epistemic-guard/blob/v0.5.2/results/table3_heldout_supervision.json\#L92}{91.0\%}} & \href{https://github.com/AlienatedAliens/epistemic-guard/blob/v0.5.2/results/table3_heldout_supervision.json\#L108}{76.67\%} & \textbf{\href{https://github.com/AlienatedAliens/epistemic-guard/blob/v0.5.2/results/table3_heldout_supervision.json\#L86}{17.9}} & \href{https://github.com/AlienatedAliens/epistemic-guard/blob/v0.5.2/results/table3_heldout_supervision.json\#L87}{35.83} & \textbf{\href{https://github.com/AlienatedAliens/epistemic-guard/blob/v0.5.2/results/table3_heldout_supervision.json\#L84}{0}} \\
\quad \emph{Ablation:} Stage-1 (no receipts) & \href{https://github.com/AlienatedAliens/epistemic-guard/blob/v0.5.2/results/table3_heldout_supervision.json\#L12}{84.8\%} [\href{https://github.com/AlienatedAliens/epistemic-guard/blob/v0.5.2/results/table3_heldout_supervision.json\#L14}{79.83\%}, \href{https://github.com/AlienatedAliens/epistemic-guard/blob/v0.5.2/results/table3_heldout_supervision.json\#L13}{88.72\%}] & \href{https://github.com/AlienatedAliens/epistemic-guard/blob/v0.5.2/results/table3_heldout_supervision.json\#L20}{97.54\%} & \href{https://github.com/AlienatedAliens/epistemic-guard/blob/v0.5.2/results/table3_heldout_supervision.json\#L16}{27.34\%} & \href{https://github.com/AlienatedAliens/epistemic-guard/blob/v0.5.2/results/table3_heldout_supervision.json\#L17}{91.0\%} & \href{https://github.com/AlienatedAliens/epistemic-guard/blob/v0.5.2/results/table3_heldout_supervision.json\#L155}{58.33\%} & \href{https://github.com/AlienatedAliens/epistemic-guard/blob/v0.5.2/results/table3_heldout_supervision.json\#L19}{24.98} & --- & \href{https://github.com/AlienatedAliens/epistemic-guard/blob/v0.5.2/results/table3_heldout_supervision.json\#L84}{0} \\
\quad \emph{Ablation:} Regex + receipts & \href{https://github.com/AlienatedAliens/epistemic-guard/blob/v0.5.2/results/table3_heldout_supervision.json\#L3}{64.0\%} [\href{https://github.com/AlienatedAliens/epistemic-guard/blob/v0.5.2/results/table3_heldout_supervision.json\#L5}{57.88\%}, \href{https://github.com/AlienatedAliens/epistemic-guard/blob/v0.5.2/results/table3_heldout_supervision.json\#L4}{69.7\%}] & \href{https://github.com/AlienatedAliens/epistemic-guard/blob/v0.5.2/results/table3_heldout_supervision.json\#L11}{30.33\%} & \href{https://github.com/AlienatedAliens/epistemic-guard/blob/v0.5.2/results/table3_heldout_supervision.json\#L7}{3.91\%} & \href{https://github.com/AlienatedAliens/epistemic-guard/blob/v0.5.2/results/table3_heldout_supervision.json\#L8}{50.0\%} & \href{https://github.com/AlienatedAliens/epistemic-guard/blob/v0.5.2/results/table3_heldout_supervision.json\#L150}{56.67\%} & \href{https://github.com/AlienatedAliens/epistemic-guard/blob/v0.5.2/results/table3_heldout_supervision.json\#L10}{0.08} & --- & \href{https://github.com/AlienatedAliens/epistemic-guard/blob/v0.5.2/results/table3_heldout_supervision.json\#L84}{0} \\
\midrule
Mistral-NeMo-12B Judge (GPU) & \href{https://github.com/AlienatedAliens/epistemic-guard/blob/v0.5.2/results/table3_heldout_supervision.json\#L161}{86.8\%} [\href{https://github.com/AlienatedAliens/epistemic-guard/blob/v0.5.2/results/table3_heldout_supervision.json\#L163}{82.04\%}, \href{https://github.com/AlienatedAliens/epistemic-guard/blob/v0.5.2/results/table3_heldout_supervision.json\#L162}{90.44\%}] & \href{https://github.com/AlienatedAliens/epistemic-guard/blob/v0.5.2/results/table3_heldout_supervision.json\#L171}{87.7\%} & \href{https://github.com/AlienatedAliens/epistemic-guard/blob/v0.5.2/results/table3_heldout_supervision.json\#L165}{14.06\%} & \href{https://github.com/AlienatedAliens/epistemic-guard/blob/v0.5.2/results/table3_heldout_supervision.json\#L169}{84.0\%} & \textbf{\href{https://github.com/AlienatedAliens/epistemic-guard/blob/v0.5.2/results/table3_heldout_supervision.json\#L137}{80.0\%}} & \href{https://github.com/AlienatedAliens/epistemic-guard/blob/v0.5.2/results/table3_heldout_supervision.json\#L167}{80.58} & \href{https://github.com/AlienatedAliens/epistemic-guard/blob/v0.5.2/results/table3_heldout_supervision.json\#L168}{102.15} & \href{https://github.com/AlienatedAliens/epistemic-guard/blob/v0.5.2/results/table3_heldout_supervision.json\#L166}{8,844} \\
Phi-4-14B Judge (GPU) & \href{https://github.com/AlienatedAliens/epistemic-guard/blob/v0.5.2/results/table3_heldout_supervision.json\#L177}{86.4\%} [\href{https://github.com/AlienatedAliens/epistemic-guard/blob/v0.5.2/results/table3_heldout_supervision.json\#L179}{81.6\%}, \href{https://github.com/AlienatedAliens/epistemic-guard/blob/v0.5.2/results/table3_heldout_supervision.json\#L178}{90.1\%}] & \textbf{\href{https://github.com/AlienatedAliens/epistemic-guard/blob/v0.5.2/results/table3_heldout_supervision.json\#L187}{100.0\%}} & \href{https://github.com/AlienatedAliens/epistemic-guard/blob/v0.5.2/results/table3_heldout_supervision.json\#L181}{26.56\%} & \href{https://github.com/AlienatedAliens/epistemic-guard/blob/v0.5.2/results/table3_heldout_supervision.json\#L185}{79.0\%} & \href{https://github.com/AlienatedAliens/epistemic-guard/blob/v0.5.2/results/table3_heldout_supervision.json\#L144}{65.0\%} & \href{https://github.com/AlienatedAliens/epistemic-guard/blob/v0.5.2/results/table3_heldout_supervision.json\#L183}{76.24} & \href{https://github.com/AlienatedAliens/epistemic-guard/blob/v0.5.2/results/table3_heldout_supervision.json\#L184}{102.25} & \href{https://github.com/AlienatedAliens/epistemic-guard/blob/v0.5.2/results/table3_heldout_supervision.json\#L182}{12,214} \\
DeBERTa-v3-xsmall NLI (CPU) & \href{https://github.com/AlienatedAliens/epistemic-guard/blob/v0.5.2/results/table3_heldout_supervision.json\#L55}{52.8\%} [\href{https://github.com/AlienatedAliens/epistemic-guard/blob/v0.5.2/results/table3_heldout_supervision.json\#L57}{46.62\%}, \href{https://github.com/AlienatedAliens/epistemic-guard/blob/v0.5.2/results/table3_heldout_supervision.json\#L56}{58.9\%}] & \href{https://github.com/AlienatedAliens/epistemic-guard/blob/v0.5.2/results/table3_heldout_supervision.json\#L68}{24.59\%} & \href{https://github.com/AlienatedAliens/epistemic-guard/blob/v0.5.2/results/table3_heldout_supervision.json\#L62}{20.31\%} & \href{https://github.com/AlienatedAliens/epistemic-guard/blob/v0.5.2/results/table3_heldout_supervision.json\#L66}{51.0\%} & \textbf{\href{https://github.com/AlienatedAliens/epistemic-guard/blob/v0.5.2/results/table3_heldout_supervision.json\#L102}{80.0\%}} & \href{https://github.com/AlienatedAliens/epistemic-guard/blob/v0.5.2/results/table3_heldout_supervision.json\#L64}{22.4} & \textbf{\href{https://github.com/AlienatedAliens/epistemic-guard/blob/v0.5.2/results/table3_heldout_supervision.json\#L65}{32.49}} & \textbf{\href{https://github.com/AlienatedAliens/epistemic-guard/blob/v0.5.2/results/table3_heldout_supervision.json\#L63}{0}} \\
\bottomrule
\end{tabular*}
\end{table*}

\paragraph{Hardware, Baselines, and Cohort Provenance.}
All systems in Table~\ref{tab:heldout} ran on one Windows\,11 workstation with an Intel Core Ultra\,9 275HX (24-core) CPU and an NVIDIA GeForce RTX\,5080 Laptop GPU (16\,GB VRAM). GPU LLM-judge baselines (\texttt{Mistral-\allowbreak Nemo-\allowbreak Instruct-\allowbreak 2407-\allowbreak Q5\_\allowbreak K\_\allowbreak M} and \texttt{Phi-4-\allowbreak Q6\_\allowbreak K}) ran via \texttt{llama.cpp} (\texttt{-ngl 99}, \texttt{-c 4096}, \texttt{temperature=0.0}) under a deterministic 3-rule auditor prompt receiving \texttt{USER\_\allowbreak PROMPT}, \texttt{VERIFIED\_\allowbreak EXECUTED\_\allowbreak TOOLS} (including raw argument strings such as \texttt{read\_file(path='config.json')}, identical to \texttt{Receipt\allowbreak Supervisor}'s input), and \texttt{ASSISTANT\_\allowbreak REPLY}, outputting \texttt{ALLOW} or \texttt{BLOCK}. Warm CPU latencies cover full \texttt{verify\_\allowbreak narration()} execution; cold-start CPU initialization (SHA-256 checks across \href{https://github.com/AlienatedAliens/epistemic-guard/blob/v0.5.2/results/table1_architecture.json\#L12}{4} \texttt{action\_\allowbreak claim} files, tokenizer load, and OpenVINO compilation) takes \href{https://github.com/AlienatedAliens/epistemic-guard/blob/v0.5.2/results/table3_heldout_supervision.json\#L78}{921.8}\,ms median (\href{https://github.com/AlienatedAliens/epistemic-guard/blob/v0.5.2/results/table3_heldout_supervision.json\#L79}{1,044.6}\,ms p95), paid once in persistent servers. The $N=\href{https://github.com/AlienatedAliens/epistemic-guard/blob/v0.5.2/results/table3_heldout_supervision.json\#L175}{250}$ frozen benchmark (\href{https://github.com/AlienatedAliens/epistemic-guard/blob/v0.5.2/results/table3_heldout_supervision.json\#L174}{122} ungrounded, \href{https://github.com/AlienatedAliens/epistemic-guard/blob/v0.5.2/results/table3_heldout_supervision.json\#L173}{128} grounded/benign; Hugging Face commit \texttt{73bbc046759e}, 2026-11-14, frozen with v0.3.0 weights) joins author-labeled speech-act cases \texttt{locked\_\allowbreak coh\_\allowbreak 50} ($n=\href{https://github.com/AlienatedAliens/epistemic-guard/blob/v0.5.2/results/table3_heldout_supervision.json\#L191}{50}$) and paraphrases \texttt{blind42\_\allowbreak coh\_\allowbreak 100} ($n=\href{https://github.com/AlienatedAliens/epistemic-guard/blob/v0.5.2/results/table3_heldout_supervision.json\#L190}{100}$) with \texttt{organic\_\allowbreak heldout\_\allowbreak h100} ($n=\href{https://github.com/AlienatedAliens/epistemic-guard/blob/v0.5.2/results/table3_heldout_supervision.json\#L192}{100}$): live tool-less chat turns (\texttt{receipts=[]}) from \texttt{gemma-4-e4b} (\href{https://github.com/AlienatedAliens/epistemic-guard/blob/v0.5.2/results/table3_heldout_supervision.json\#L48}{50}) and \texttt{gemma-4-26b-a4b} (\href{https://github.com/AlienatedAliens/epistemic-guard/blob/v0.5.2/results/table3_heldout_supervision.json\#L47}{50}), labeled by blinded LLM raters (\href{https://github.com/AlienatedAliens/epistemic-guard/blob/v0.5.2/results/table3_heldout_supervision.json\#L44}{30}/\href{https://github.com/AlienatedAliens/epistemic-guard/blob/v0.5.2/results/table3_heldout_supervision.json\#L46}{30}, \href{https://github.com/AlienatedAliens/epistemic-guard/blob/v0.5.2/results/table3_heldout_supervision.json\#L45}{100.0\%} dual-slice agreement) and author audit~\cite{maxwell2026evasion}. Because \texttt{organic\_\allowbreak heldout\_\allowbreak h100} is tool-less, we also generated a post-release end-to-end cohort \texttt{live\_\allowbreak mcp\_\allowbreak e2e\_\allowbreak 60} ($N=\href{https://github.com/AlienatedAliens/epistemic-guard/blob/v0.5.2/results/table3_heldout_supervision.json\#L136}{60}$: \href{https://github.com/AlienatedAliens/epistemic-guard/blob/v0.5.2/results/table3_heldout_supervision.json\#L124}{30} \texttt{gemma-4-E4B} and \href{https://github.com/AlienatedAliens/epistemic-guard/blob/v0.5.2/results/table3_heldout_supervision.json\#L143}{30} \texttt{Nemotron-3.5-30B-A3B} turns, \href{https://github.com/AlienatedAliens/epistemic-guard/blob/v0.5.2/results/table3_heldout_supervision.json\#L135}{26} ungrounded and \href{https://github.com/AlienatedAliens/epistemic-guard/blob/v0.5.2/results/table3_heldout_supervision.json\#L134}{34} grounded, labeled against wire state prior to evaluation) routed through live \texttt{mcp-proxy} child servers across \texttt{ok=true} completion, target-drift \texttt{ok=true} calls, \texttt{isError=true}, JSON-RPC \texttt{-32602}, mid-call crash, and compound two-tool turns.

\paragraph{Accuracy, Ablations, and Live MCP Receipt Binding.}
On the $N=\href{https://github.com/AlienatedAliens/epistemic-guard/blob/v0.5.2/results/table3_heldout_supervision.json\#L175}{250}$ frozen benchmark, \texttt{epistemic-guard} reaches \href{https://github.com/AlienatedAliens/epistemic-guard/blob/v0.5.2/results/table3_heldout_supervision.json\#L71}{93.2\%} accuracy (\href{https://github.com/AlienatedAliens/epistemic-guard/blob/v0.5.2/results/table3_heldout_supervision.json\#L80}{233}/\href{https://github.com/AlienatedAliens/epistemic-guard/blob/v0.5.2/results/table3_heldout_supervision.json\#L175}{250}; $\mathrm{TP}=\href{https://github.com/AlienatedAliens/epistemic-guard/blob/v0.5.2/results/table3_heldout_supervision.json\#L99}{118}$, $\mathrm{TN}=\href{https://github.com/AlienatedAliens/epistemic-guard/blob/v0.5.2/results/table3_heldout_supervision.json\#L98}{115}$, $\mathrm{FP}=\href{https://github.com/AlienatedAliens/epistemic-guard/blob/v0.5.2/results/table3_heldout_supervision.json\#L82}{13}$, $\mathrm{FN}=\href{https://github.com/AlienatedAliens/epistemic-guard/blob/v0.5.2/results/table3_heldout_supervision.json\#L81}{4}$): \href{https://github.com/AlienatedAliens/epistemic-guard/blob/v0.5.2/results/table3_heldout_supervision.json\#L88}{100.0\%} (\href{https://github.com/AlienatedAliens/epistemic-guard/blob/v0.5.2/results/table3_heldout_supervision.json\#L91}{50}/\href{https://github.com/AlienatedAliens/epistemic-guard/blob/v0.5.2/results/table3_heldout_supervision.json\#L191}{50}, CI [\href{https://github.com/AlienatedAliens/epistemic-guard/blob/v0.5.2/results/table3_heldout_supervision.json\#L90}{92.87\%}, \href{https://github.com/AlienatedAliens/epistemic-guard/blob/v0.5.2/results/table3_heldout_supervision.json\#L89}{100.0\%}]) on \texttt{locked\_\allowbreak coh\_\allowbreak 50}, \href{https://github.com/AlienatedAliens/epistemic-guard/blob/v0.5.2/results/table3_heldout_supervision.json\#L72}{92.0\%} (\href{https://github.com/AlienatedAliens/epistemic-guard/blob/v0.5.2/results/table3_heldout_supervision.json\#L75}{92}/\href{https://github.com/AlienatedAliens/epistemic-guard/blob/v0.5.2/results/table3_heldout_supervision.json\#L190}{100}, CI [\href{https://github.com/AlienatedAliens/epistemic-guard/blob/v0.5.2/results/table3_heldout_supervision.json\#L74}{85.0\%}, \href{https://github.com/AlienatedAliens/epistemic-guard/blob/v0.5.2/results/table3_heldout_supervision.json\#L73}{95.89\%}]) on \texttt{blind42\_\allowbreak coh\_\allowbreak 100}, and \href{https://github.com/AlienatedAliens/epistemic-guard/blob/v0.5.2/results/table3_heldout_supervision.json\#L92}{91.0\%} (\href{https://github.com/AlienatedAliens/epistemic-guard/blob/v0.5.2/results/table3_heldout_supervision.json\#L95}{91}/\href{https://github.com/AlienatedAliens/epistemic-guard/blob/v0.5.2/results/table3_heldout_supervision.json\#L192}{100}, CI [\href{https://github.com/AlienatedAliens/epistemic-guard/blob/v0.5.2/results/table3_heldout_supervision.json\#L94}{83.77\%}, \href{https://github.com/AlienatedAliens/epistemic-guard/blob/v0.5.2/results/table3_heldout_supervision.json\#L93}{95.19\%}]) on \texttt{organic\_\allowbreak heldout\_\allowbreak h100} at \href{https://github.com/AlienatedAliens/epistemic-guard/blob/v0.5.2/results/table3_heldout_supervision.json\#L86}{17.9}\,ms median turn latency (\href{https://github.com/AlienatedAliens/epistemic-guard/blob/v0.5.2/results/table3_heldout_supervision.json\#L97}{8.62}\,ms single-clause, \href{https://github.com/AlienatedAliens/epistemic-guard/blob/v0.5.2/results/table3_heldout_supervision.json\#L85}{19.38}\,ms mean, \href{https://github.com/AlienatedAliens/epistemic-guard/blob/v0.5.2/results/table3_heldout_supervision.json\#L87}{35.83}\,ms p95). At the v0.3.0 freeze it scored \href{https://github.com/AlienatedAliens/epistemic-guard/blob/v0.5.2/results/table3_heldout_supervision.json\#L50}{94.4\%} (\href{https://github.com/AlienatedAliens/epistemic-guard/blob/v0.5.2/results/table3_heldout_supervision.json\#L51}{236}/\href{https://github.com/AlienatedAliens/epistemic-guard/blob/v0.5.2/results/table3_heldout_supervision.json\#L175}{250}; \href{https://github.com/AlienatedAliens/epistemic-guard/blob/v0.5.2/results/table3_heldout_supervision.json\#L49}{95}/\href{https://github.com/AlienatedAliens/epistemic-guard/blob/v0.5.2/results/table3_heldout_supervision.json\#L190}{100} on \texttt{blind42\_\allowbreak coh\_\allowbreak 100}); v0.5.1--v0.5.2 security hardening (rejecting zero-argument wildcard matches on named files) cost \href{https://github.com/AlienatedAliens/epistemic-guard/blob/v0.5.2/results/table3_heldout_supervision.json\#L52}{3} bare-tool items, so symbolic rules are audit-hardened rather than blind while neural weights stayed frozen. Regex+receipts scores \href{https://github.com/AlienatedAliens/epistemic-guard/blob/v0.5.2/results/table3_heldout_supervision.json\#L3}{64.0\%} (\href{https://github.com/AlienatedAliens/epistemic-guard/blob/v0.5.2/results/table3_heldout_supervision.json\#L6}{160}/\href{https://github.com/AlienatedAliens/epistemic-guard/blob/v0.5.2/results/table3_heldout_supervision.json\#L175}{250}; \href{https://github.com/AlienatedAliens/epistemic-guard/blob/v0.5.2/results/table3_heldout_supervision.json\#L11}{30.33\%} recall), and Stage-1-only (\texttt{receipts=[]}) scores \href{https://github.com/AlienatedAliens/epistemic-guard/blob/v0.5.2/results/table3_heldout_supervision.json\#L12}{84.8\%} (\href{https://github.com/AlienatedAliens/epistemic-guard/blob/v0.5.2/results/table3_heldout_supervision.json\#L15}{212}/\href{https://github.com/AlienatedAliens/epistemic-guard/blob/v0.5.2/results/table3_heldout_supervision.json\#L175}{250}; \href{https://github.com/AlienatedAliens/epistemic-guard/blob/v0.5.2/results/table3_heldout_supervision.json\#L16}{27.34\%} FPR). Stage-1-only exhibits a higher median latency (\href{https://github.com/AlienatedAliens/epistemic-guard/blob/v0.5.2/results/table3_heldout_supervision.json\#L19}{24.98}\,ms vs.\ \href{https://github.com/AlienatedAliens/epistemic-guard/blob/v0.5.2/results/table3_heldout_supervision.json\#L86}{17.9}\,ms) because \texttt{Receipt\allowbreak Supervisor} short-circuits conversational clauses lacking action/state signals on multi-clause turns when \texttt{ok\_receipts} is non-empty, whereas passing \texttt{receipts=[]} runs OpenVINO on every clause.

Overall ($N=\href{https://github.com/AlienatedAliens/epistemic-guard/blob/v0.5.2/results/table3_heldout_supervision.json\#L175}{250}$), \texttt{epistemic-guard} beats Mistral-NeMo-12B by $+\href{https://github.com/AlienatedAliens/epistemic-guard/blob/v0.5.2/results/table3_heldout_supervision.json\#L24}{6.4}$\,pp (\href{https://github.com/AlienatedAliens/epistemic-guard/blob/v0.5.2/results/table3_heldout_supervision.json\#L161}{86.8\%}, \href{https://github.com/AlienatedAliens/epistemic-guard/blob/v0.5.2/results/table3_heldout_supervision.json\#L164}{217}/\href{https://github.com/AlienatedAliens/epistemic-guard/blob/v0.5.2/results/table3_heldout_supervision.json\#L175}{250}; McNemar \href{https://github.com/AlienatedAliens/epistemic-guard/blob/v0.5.2/results/table3_heldout_supervision.json\#L35}{31} vs.\ \href{https://github.com/AlienatedAliens/epistemic-guard/blob/v0.5.2/results/table3_heldout_supervision.json\#L34}{15}, $p=\href{https://github.com/AlienatedAliens/epistemic-guard/blob/v0.5.2/results/table3_heldout_supervision.json\#L36}{0.0259}$; $+\href{https://github.com/AlienatedAliens/epistemic-guard/blob/v0.5.2/results/table3_heldout_supervision.json\#L26}{7.0}$\,pp on \texttt{Org-100}, \href{https://github.com/AlienatedAliens/epistemic-guard/blob/v0.5.2/results/table3_heldout_supervision.json\#L29}{14} vs.\ \href{https://github.com/AlienatedAliens/epistemic-guard/blob/v0.5.2/results/table3_heldout_supervision.json\#L28}{7}, $p=\href{https://github.com/AlienatedAliens/epistemic-guard/blob/v0.5.2/results/table3_heldout_supervision.json\#L30}{0.1892}$; \href{https://github.com/AlienatedAliens/epistemic-guard/blob/v0.5.2/results/table3_heldout_supervision.json\#L40}{4.5}$\times$ faster p50, saving \href{https://github.com/AlienatedAliens/epistemic-guard/blob/v0.5.2/results/table3_heldout_supervision.json\#L166}{8,844}\,MiB VRAM) and Phi-4-14B by $+\href{https://github.com/AlienatedAliens/epistemic-guard/blob/v0.5.2/results/table3_heldout_supervision.json\#L25}{6.8}$\,pp (\href{https://github.com/AlienatedAliens/epistemic-guard/blob/v0.5.2/results/table3_heldout_supervision.json\#L177}{86.4\%}, \href{https://github.com/AlienatedAliens/epistemic-guard/blob/v0.5.2/results/table3_heldout_supervision.json\#L180}{216}/\href{https://github.com/AlienatedAliens/epistemic-guard/blob/v0.5.2/results/table3_heldout_supervision.json\#L175}{250}; McNemar \href{https://github.com/AlienatedAliens/epistemic-guard/blob/v0.5.2/results/table3_heldout_supervision.json\#L38}{28} vs.\ \href{https://github.com/AlienatedAliens/epistemic-guard/blob/v0.5.2/results/table3_heldout_supervision.json\#L37}{11}, $p=\href{https://github.com/AlienatedAliens/epistemic-guard/blob/v0.5.2/results/table3_heldout_supervision.json\#L39}{0.0095}$; $+\href{https://github.com/AlienatedAliens/epistemic-guard/blob/v0.5.2/results/table3_heldout_supervision.json\#L27}{12.0}$\,pp on \texttt{Org-100}, \href{https://github.com/AlienatedAliens/epistemic-guard/blob/v0.5.2/results/table3_heldout_supervision.json\#L32}{17} vs.\ \href{https://github.com/AlienatedAliens/epistemic-guard/blob/v0.5.2/results/table3_heldout_supervision.json\#L31}{5}, $p=\href{https://github.com/AlienatedAliens/epistemic-guard/blob/v0.5.2/results/table3_heldout_supervision.json\#L33}{0.0169}$; \href{https://github.com/AlienatedAliens/epistemic-guard/blob/v0.5.2/results/table3_heldout_supervision.json\#L41}{4.26}$\times$ faster p50, saving \href{https://github.com/AlienatedAliens/epistemic-guard/blob/v0.5.2/results/table3_heldout_supervision.json\#L182}{12,214}\,MiB VRAM), while DeBERTa-v3 NLI scores \href{https://github.com/AlienatedAliens/epistemic-guard/blob/v0.5.2/results/table3_heldout_supervision.json\#L55}{52.8\%} (\href{https://github.com/AlienatedAliens/epistemic-guard/blob/v0.5.2/results/table3_heldout_supervision.json\#L58}{132}/\href{https://github.com/AlienatedAliens/epistemic-guard/blob/v0.5.2/results/table3_heldout_supervision.json\#L175}{250}, $+\href{https://github.com/AlienatedAliens/epistemic-guard/blob/v0.5.2/results/table3_heldout_supervision.json\#L23}{40.4}$\,pp gap; \href{https://github.com/AlienatedAliens/epistemic-guard/blob/v0.5.2/results/table3_heldout_supervision.json\#L60}{73}/\href{https://github.com/AlienatedAliens/epistemic-guard/blob/v0.5.2/results/table3_heldout_supervision.json\#L59}{92} false negatives, \href{https://github.com/AlienatedAliens/epistemic-guard/blob/v0.5.2/results/table3_heldout_supervision.json\#L61}{79.35\%}, predict \emph{neutral}). On \texttt{Live-60} (zero-shot post-release, no custom tool stems), the supervisor deliberately favors recall over false alarms---a false positive costs one \texttt{to\_retry\_prompt()} retry whereas a missed fabrication corrupts state---catching \href{https://github.com/AlienatedAliens/epistemic-guard/blob/v0.5.2/results/table3_heldout_supervision.json\#L121}{96.15\%} of hazards (\href{https://github.com/AlienatedAliens/epistemic-guard/blob/v0.5.2/results/table3_heldout_supervision.json\#L123}{25}/\href{https://github.com/AlienatedAliens/epistemic-guard/blob/v0.5.2/results/table3_heldout_supervision.json\#L135}{26}, CI [\href{https://github.com/AlienatedAliens/epistemic-guard/blob/v0.5.2/results/table3_heldout_supervision.json\#L120}{81.11\%}, \href{https://github.com/AlienatedAliens/epistemic-guard/blob/v0.5.2/results/table3_heldout_supervision.json\#L119}{99.32\%}], \href{https://github.com/AlienatedAliens/epistemic-guard/blob/v0.5.2/results/table3_heldout_supervision.json\#L112}{1} missed retry promise) at \href{https://github.com/AlienatedAliens/epistemic-guard/blob/v0.5.2/results/table3_heldout_supervision.json\#L116}{38.24\%} FPR (\href{https://github.com/AlienatedAliens/epistemic-guard/blob/v0.5.2/results/table3_heldout_supervision.json\#L113}{13}/\href{https://github.com/AlienatedAliens/epistemic-guard/blob/v0.5.2/results/table3_heldout_supervision.json\#L134}{34}) and \href{https://github.com/AlienatedAliens/epistemic-guard/blob/v0.5.2/results/table3_heldout_supervision.json\#L108}{76.67\%} accuracy (\href{https://github.com/AlienatedAliens/epistemic-guard/blob/v0.5.2/results/table3_heldout_supervision.json\#L111}{46}/\href{https://github.com/AlienatedAliens/epistemic-guard/blob/v0.5.2/results/table3_heldout_supervision.json\#L136}{60}, CI [\href{https://github.com/AlienatedAliens/epistemic-guard/blob/v0.5.2/results/table3_heldout_supervision.json\#L110}{64.56\%}, \href{https://github.com/AlienatedAliens/epistemic-guard/blob/v0.5.2/results/table3_heldout_supervision.json\#L109}{85.56\%}]) in \href{https://github.com/AlienatedAliens/epistemic-guard/blob/v0.5.2/results/table3_heldout_supervision.json\#L117}{9.16}\,ms p50 (\href{https://github.com/AlienatedAliens/epistemic-guard/blob/v0.5.2/results/table3_heldout_supervision.json\#L118}{23.12}\,ms p95). Stage-2 receipt binding adds $+\href{https://github.com/AlienatedAliens/epistemic-guard/blob/v0.5.2/results/table3_heldout_supervision.json\#L107}{18.34}$\,pp over Stage-1-only (\href{https://github.com/AlienatedAliens/epistemic-guard/blob/v0.5.2/results/table3_heldout_supervision.json\#L155}{58.33\%}, \href{https://github.com/AlienatedAliens/epistemic-guard/blob/v0.5.2/results/table3_heldout_supervision.json\#L156}{35}/\href{https://github.com/AlienatedAliens/epistemic-guard/blob/v0.5.2/results/table3_heldout_supervision.json\#L136}{60}, \href{https://github.com/AlienatedAliens/epistemic-guard/blob/v0.5.2/results/table3_heldout_supervision.json\#L157}{70.59\%} FPR; \href{https://github.com/AlienatedAliens/epistemic-guard/blob/v0.5.2/results/table3_heldout_supervision.json\#L132}{11} vs.\ \href{https://github.com/AlienatedAliens/epistemic-guard/blob/v0.5.2/results/table3_heldout_supervision.json\#L131}{0}, $p=\href{https://github.com/AlienatedAliens/epistemic-guard/blob/v0.5.2/results/table3_heldout_supervision.json\#L133}{0.001}$), while Regex+receipts has \href{https://github.com/AlienatedAliens/epistemic-guard/blob/v0.5.2/results/table3_heldout_supervision.json\#L152}{0.0\%} recall (\href{https://github.com/AlienatedAliens/epistemic-guard/blob/v0.5.2/results/table3_heldout_supervision.json\#L151}{34}/\href{https://github.com/AlienatedAliens/epistemic-guard/blob/v0.5.2/results/table3_heldout_supervision.json\#L136}{60}) and DeBERTa-v3 NLI shows why accuracy misleads on uneven classes (\href{https://github.com/AlienatedAliens/epistemic-guard/blob/v0.5.2/results/table3_heldout_supervision.json\#L102}{80.0\%}, \href{https://github.com/AlienatedAliens/epistemic-guard/blob/v0.5.2/results/table3_heldout_supervision.json\#L103}{48}/\href{https://github.com/AlienatedAliens/epistemic-guard/blob/v0.5.2/results/table3_heldout_supervision.json\#L136}{60}, at \href{https://github.com/AlienatedAliens/epistemic-guard/blob/v0.5.2/results/table3_heldout_supervision.json\#L105}{53.85\%} recall, \href{https://github.com/AlienatedAliens/epistemic-guard/blob/v0.5.2/results/table3_heldout_supervision.json\#L104}{12}/\href{https://github.com/AlienatedAliens/epistemic-guard/blob/v0.5.2/results/table3_heldout_supervision.json\#L135}{26} missed). Live accuracy gaps vs.\ GPU judges are non-significant---\texttt{epistemic-guard} is nominally $+\href{https://github.com/AlienatedAliens/epistemic-guard/blob/v0.5.2/results/table3_heldout_supervision.json\#L106}{11.67}$\,pp above Phi-4-14B (\href{https://github.com/AlienatedAliens/epistemic-guard/blob/v0.5.2/results/table3_heldout_supervision.json\#L144}{65.0\%}, \href{https://github.com/AlienatedAliens/epistemic-guard/blob/v0.5.2/results/table3_heldout_supervision.json\#L145}{39}/\href{https://github.com/AlienatedAliens/epistemic-guard/blob/v0.5.2/results/table3_heldout_supervision.json\#L136}{60}, \href{https://github.com/AlienatedAliens/epistemic-guard/blob/v0.5.2/results/table3_heldout_supervision.json\#L147}{61.76\%} FPR; \href{https://github.com/AlienatedAliens/epistemic-guard/blob/v0.5.2/results/table3_heldout_supervision.json\#L129}{13} vs.\ \href{https://github.com/AlienatedAliens/epistemic-guard/blob/v0.5.2/results/table3_heldout_supervision.json\#L128}{6}, $p=\href{https://github.com/AlienatedAliens/epistemic-guard/blob/v0.5.2/results/table3_heldout_supervision.json\#L130}{0.1671}$; \href{https://github.com/AlienatedAliens/epistemic-guard/blob/v0.5.2/results/table3_heldout_supervision.json\#L154}{7.32}$\times$ faster vs.\ \href{https://github.com/AlienatedAliens/epistemic-guard/blob/v0.5.2/results/table3_heldout_supervision.json\#L148}{67.04}\,ms) and nominally $\href{https://github.com/AlienatedAliens/epistemic-guard/blob/v0.5.2/results/table3_heldout_supervision.json\#L140}{3.33}$\,pp below Mistral-NeMo-12B (\href{https://github.com/AlienatedAliens/epistemic-guard/blob/v0.5.2/results/table3_heldout_supervision.json\#L137}{80.0\%}, \href{https://github.com/AlienatedAliens/epistemic-guard/blob/v0.5.2/results/table3_heldout_supervision.json\#L138}{48}/\href{https://github.com/AlienatedAliens/epistemic-guard/blob/v0.5.2/results/table3_heldout_supervision.json\#L136}{60}, \href{https://github.com/AlienatedAliens/epistemic-guard/blob/v0.5.2/results/table3_heldout_supervision.json\#L139}{35.29\%} FPR; \href{https://github.com/AlienatedAliens/epistemic-guard/blob/v0.5.2/results/table3_heldout_supervision.json\#L126}{6} vs.\ \href{https://github.com/AlienatedAliens/epistemic-guard/blob/v0.5.2/results/table3_heldout_supervision.json\#L125}{8}, $p=\href{https://github.com/AlienatedAliens/epistemic-guard/blob/v0.5.2/results/table3_heldout_supervision.json\#L127}{0.7905}$) at \href{https://github.com/AlienatedAliens/epistemic-guard/blob/v0.5.2/results/table3_heldout_supervision.json\#L153}{7.83}$\times$ lower p50 (\href{https://github.com/AlienatedAliens/epistemic-guard/blob/v0.5.2/results/table3_heldout_supervision.json\#L117}{9.16} vs.\ \href{https://github.com/AlienatedAliens/epistemic-guard/blob/v0.5.2/results/table3_heldout_supervision.json\#L141}{71.69}\,ms) and \href{https://github.com/AlienatedAliens/epistemic-guard/blob/v0.5.2/results/table3_heldout_supervision.json\#L84}{0}\,MiB VRAM; its \href{https://github.com/AlienatedAliens/epistemic-guard/blob/v0.5.2/results/table3_heldout_supervision.json\#L113}{13} FPs comprise \href{https://github.com/AlienatedAliens/epistemic-guard/blob/v0.5.2/results/table3_heldout_supervision.json\#L115}{6} custom tool names (\texttt{execute\_sql}, \texttt{http\_post}, \texttt{git\_commit}) and \href{https://github.com/AlienatedAliens/epistemic-guard/blob/v0.5.2/results/table3_heldout_supervision.json\#L114}{7} comma-split \texttt{"attempted X, but failed"} clauses.

\section{Limitations and Threats to Validity}
\label{sec:limitations}

\paragraph{Internal Validity.}
Protocol conformance workloads in Table~\ref{tab:conformance} run real child processes, OS file locks, Ed25519/HKDF signing, and OpenVINO INT8 inference on deterministic local payloads rather than live SaaS endpoints so auditors can run \texttt{osaudit results -{-}rerun} offline without API keys or network variance. Supervision labels derive from author rubrics, blinded LLM raters plus author audit, and pre-committed MCP wire-state rules rather than multi-human inter-annotator ($\kappa$) studies, and commercial frontier API judges were omitted to keep evaluation zero-API-cost and offline-reproducible.

\paragraph{External Validity.}
Our supervision cohorts ($N=\href{https://github.com/AlienatedAliens/epistemic-guard/blob/v0.5.2/results/table3_heldout_supervision.json\#L175}{250}$ frozen, $N=\href{https://github.com/AlienatedAliens/epistemic-guard/blob/v0.5.2/results/table3_heldout_supervision.json\#L136}{60}$ live MCP) evaluate English replies on one workstation; as \texttt{Live-60} shows (\href{https://github.com/AlienatedAliens/epistemic-guard/blob/v0.5.2/results/table3_heldout_supervision.json\#L115}{6} unmapped-stem and \href{https://github.com/AlienatedAliens/epistemic-guard/blob/v0.5.2/results/table3_heldout_supervision.json\#L114}{7} cross-clause failure-report FPs), custom MCP tool vocabularies and multi-clause \texttt{"attempted X, but failed"} reports raise false positives until custom tool stems and cross-clause negation propagation are configured.

\paragraph{Construct and Transport Scope.}
First, \texttt{mcp-proxy} in v0.5.2 mediates \texttt{stdio} MCP servers; Streamable HTTP/SSE remote servers require a local \texttt{stdio} bridge or in-process \texttt{ReceiptRecorder} wrapping. Second, receipts prove what a downstream server returned on the wire; a buggy or malicious server returning forged \texttt{ok=true} payloads yields a validly signed false receipt, and custom servers rejecting calls via non-standard error codes are conservatively logged as \texttt{executed=true, ok=false}. Third, as quantified in Section~\ref{sec:rq1}, $k=\href{https://github.com/AlienatedAliens/epistemic-guard/blob/v0.5.2/results/table1_architecture.json\#L29}{4}$ shingle/SimHash clustering collapses verbatim and lightly edited wire copy (\href{https://github.com/AlienatedAliens/epistemic-guard/blob/v0.5.2/results/table2_conformance.json\#L77}{100.0\%}) but does not collapse moderate or abstractive rewrites (\href{https://github.com/AlienatedAliens/epistemic-guard/blob/v0.5.2/results/table2_conformance.json\#L78}{0.0\%}).

\section{Related Work}
\label{sec:related}

\paragraph{MCP Firewalls and Cryptographic Agent Runtimes.}
Pipelock~\cite{pipelock2026} is an open-source Go egress proxy and MCP firewall that scans outbound tool parameters for prompt injection or data leaks and emits Ed25519 receipts in \texttt{AER-1} format~\cite{zambo2026aer1}. NabaOS~\cite{nabaos2026} runs agents inside OS sandboxes governed by Ed25519-signed YAML constitutions. Pipelock inspects traffic before execution; \texttt{epistemic-guard} imports and exports \texttt{AER-1} receipts while checking post-turn prose (\texttt{ReceiptSupervisor}), gating SQLite memory writes, and collapsing syndicated RAG sources.

\paragraph{Conversational Guardrails and LLM-as-a-Judge.}
NeMo Guardrails~\cite{rebedea2023nemo} executes Colang dialogue rails; Llama Guard~\cite{inan2023llamaguard} fine-tunes multi-billion-parameter classifiers for safety moderation; and Guardrails AI~\cite{guardrailsai2024} chains structural and semantic validators. Rather than moderating toxicity or running an \href{https://github.com/AlienatedAliens/epistemic-guard/blob/v0.5.2/results/table3_heldout_supervision.json\#L166}{8,844}--\href{https://github.com/AlienatedAliens/epistemic-guard/blob/v0.5.2/results/table3_heldout_supervision.json\#L182}{12,214}\,MiB GPU judge, \texttt{epistemic-guard} checks whether user-facing prose matches signed wire receipts in \href{https://github.com/AlienatedAliens/epistemic-guard/blob/v0.5.2/results/table3_heldout_supervision.json\#L84}{0}\,MiB VRAM.

\paragraph{Classical Runtime Verification and Provenance Security.}
Runtime verification monitors execution traces against formal properties~\cite{leucker2009brief,bartocci2018lectures}, and Certificate Transparency~\cite{laurie2013rfc6962} introduced append-only Merkle audit logs for untrusted intermediaries. Work on indirect prompt injection~\cite{greshake2023indirect} and agent memory poisoning~\cite{maxwell2026poisoning,maxwell2026fivelayer} shows that mixing untrusted text and verified execution state in one memory store enables cross-session compromise (\texttt{ASI06}~\cite{owasp2026agentic}). \texttt{epistemic-guard} applies out-of-process monitor separation and RFC\,6962 Merkle logs directly to Python tools and \texttt{stdio} MCP servers.

\section{Conclusion}
\label{sec:conclusion}

\texttt{epistemic-guard} (v0.5.2) combines an out-of-process \texttt{stdio} MCP receipt proxy, OpenVINO INT8 prose supervision, Ed25519-gated SQLite storage, and wire-syndication deduplication. Across $N=\href{https://github.com/AlienatedAliens/epistemic-guard/blob/v0.5.2/results/table3_heldout_supervision.json\#L175}{250}$ frozen-benchmark traces and $N=\href{https://github.com/AlienatedAliens/epistemic-guard/blob/v0.5.2/results/table3_heldout_supervision.json\#L136}{60}$ live \texttt{mcp-proxy} fault-injection traces, \texttt{ReceiptSupervisor} reaches \href{https://github.com/AlienatedAliens/epistemic-guard/blob/v0.5.2/results/table3_heldout_supervision.json\#L71}{93.2\%} and \href{https://github.com/AlienatedAliens/epistemic-guard/blob/v0.5.2/results/table3_heldout_supervision.json\#L108}{76.67\%} accuracy (\href{https://github.com/AlienatedAliens/epistemic-guard/blob/v0.5.2/results/table3_heldout_supervision.json\#L96}{96.72\%} and \href{https://github.com/AlienatedAliens/epistemic-guard/blob/v0.5.2/results/table3_heldout_supervision.json\#L121}{96.15\%} recall; \href{https://github.com/AlienatedAliens/epistemic-guard/blob/v0.5.2/results/table3_heldout_supervision.json\#L83}{10.16\%} and \href{https://github.com/AlienatedAliens/epistemic-guard/blob/v0.5.2/results/table3_heldout_supervision.json\#L116}{38.24\%} FPR; $+\href{https://github.com/AlienatedAliens/epistemic-guard/blob/v0.5.2/results/table3_heldout_supervision.json\#L107}{18.34}$\,pp over Stage-1-only on live MCP traffic, $p=\href{https://github.com/AlienatedAliens/epistemic-guard/blob/v0.5.2/results/table3_heldout_supervision.json\#L133}{0.001}$) in \href{https://github.com/AlienatedAliens/epistemic-guard/blob/v0.5.2/results/table3_heldout_supervision.json\#L117}{9.16}--\href{https://github.com/AlienatedAliens/epistemic-guard/blob/v0.5.2/results/table3_heldout_supervision.json\#L86}{17.9}\,ms median CPU latency and \href{https://github.com/AlienatedAliens/epistemic-guard/blob/v0.5.2/results/table3_heldout_supervision.json\#L84}{0}\,MiB VRAM, while \texttt{SQLiteProvenanceStore} and \texttt{mcp-proxy} block \href{https://github.com/AlienatedAliens/epistemic-guard/blob/v0.5.2/results/table2_conformance.json\#L60}{100.0\%} of unauthorized writes and tamper/replay faults. Future work targets user-configurable MCP tool-stem registration, cross-clause failure-negation propagation, Streamable HTTP proxying, and lightweight paraphrase clustering.

\paragraph{Data Availability, Acknowledgments, and Disclosures.}
{\sloppy Source code is MIT-licensed at \url{https://github.com/AlienatedAliens/epistemic-guard}; INT8 model weights (\texttt{Apache-2.0}/\texttt{MIT} upstream) and benchmark corpora (with SHA-256 digests and OpenTimestamps receipts on web, hazard, and NLI splits) are archived on Hugging Face (\url{https://huggingface.co/RollSolo}). We thank Martin Monperrus for \texttt{osaudit}~\cite{monperrus2026osaudit} and Brennan Zambo for the open \texttt{AER-1} conformance corpus~\cite{zambo2026aer1}. Sean Maxwell is founder of Roll Solo (Member of NVIDIA Inception); no external funding was received.\par}

{\small\raggedright
\bibliographystyle{plain}
\bibliography{references}

\begin{thebibliography}{10}

\bibitem{bartocci2018lectures}
Ezio Bartocci and Yli{\`e}s Falcone, editors.
\newblock {\em Lectures on Runtime Verification: Introductory and Advanced
  Topics}, volume 10457 of {\em Lecture Notes in Computer Science}.
\newblock Springer, 2018.

\bibitem{broder1997resemblance}
Andrei~Z. Broder.
\newblock On the resemblance and containment of documents.
\newblock In {\em Proceedings of Compression and Complexity of SEQUENCES 1997},
  pages 21--29. IEEE, 1997.

\bibitem{charikar2002similarity}
Moses~S. Charikar.
\newblock Similarity estimation techniques from rounding algorithms.
\newblock In {\em Proceedings of the Thirty-Fourth Annual ACM Symposium on
  Theory of Computing}, pages 380--388. ACM, 2002.

\bibitem{greshake2023indirect}
Kai Greshake, Sahar Abdelnabi, Shailesh Mishra, Christoph Endres, Thorsten
  Holz, and Mario Fritz.
\newblock Not what you've signed up for: Compromising real-world
  {LLM}-integrated applications with indirect prompt injection.
\newblock In {\em Proceedings of the 16th ACM Workshop on Artificial
  Intelligence and Security}, pages 79--90. ACM, 2023.

\bibitem{guardrailsai2024}
{Guardrails AI Contributors}.
\newblock {Guardrails AI}: Adding guardrails to large language models.
\newblock \url{https://github.com/guardrails-ai/guardrails}, 2024.
\newblock Accessed: 2026-11-19.

\bibitem{he2023debertav3}
Pengcheng He, Jianfeng Gao, and Weizhu Chen.
\newblock {DeBERTaV3}: Improving {DeBERTa} using {ELECTRA}-style pre-training
  with gradient-disentangled embedding sharing.
\newblock In {\em International Conference on Learning Representations}, 2023.

\bibitem{inan2023llamaguard}
Hakan Inan, Kartikeya Upasani, Jianfeng Chi, Rashi Rungta, Krithika Iyer,
  Yuning Mao, Michael Tontchev, Qing Hu, Brian Fuller, Davide Testuggine, and
  Madian Khabsa.
\newblock {Llama Guard}: {LLM}-based input-output safeguard for human-{AI}
  conversations.
\newblock arXiv:2312.06674, \url{https://arxiv.org/abs/2312.06674}, 2023.

\bibitem{josefsson2017rfc8032}
Simon Josefsson and Ilari Liusvaara.
\newblock Edwards-curve digital signature algorithm ({EdDSA}).
\newblock RFC 8032, Internet Engineering Task Force, 2017.

\bibitem{krawczyk2010rfc5869}
Hugo Krawczyk and Pasi Eronen.
\newblock {HMAC}-based extract-and-expand key derivation function ({HKDF}).
\newblock RFC 5869, Internet Engineering Task Force, 2010.

\bibitem{laurie2013rfc6962}
Ben Laurie, Adam Langley, and Emilia Kasper.
\newblock {Certificate Transparency}.
\newblock RFC 6962, Internet Engineering Task Force, 2013.

\bibitem{leucker2009brief}
Martin Leucker and Christian Schallhart.
\newblock A brief account of runtime verification.
\newblock {\em The Journal of Logic and Algebraic Programming}, 78(5):293--303,
  2009.

\bibitem{maxwell2026fivelayer}
Sean Maxwell.
\newblock Closing the epistemic loop across the five layers of agent memory:
  Cryptographic provenance, paraconsistent bilattices, and {Z3}-quarantined
  procedural induction.
\newblock Zenodo preprint, \url{https://doi.org/10.5281/zenodo.23901683}, 2026.

\bibitem{maxwell2026poisoning}
Sean Maxwell.
\newblock Cryptographic provenance gates mitigate multi-session memory
  poisoning in autonomous agents.
\newblock Zenodo preprint, \url{https://doi.org/10.5281/zenodo.23901452}, 2026.

\bibitem{maxwell2026seed}
Sean Maxwell.
\newblock Provenance-gated memory and deontic output supervision for
  epistemically grounded autonomous agents.
\newblock Zenodo preprint, \url{https://doi.org/10.5281/zenodo.23649049}, 2026.

\bibitem{maxwell2026evasion}
Sean Maxwell.
\newblock Why speech-regex action supervisors miss live output: An empirical
  evasion taxonomy and held-out structured action contract evaluation.
\newblock Zenodo preprint, \url{https://doi.org/10.5281/zenodo.23901605}, 2026.

\bibitem{mcnemar1947note}
Quinn McNemar.
\newblock Note on the sampling error of the difference between correlated
  proportions or percentages.
\newblock {\em Psychometrika}, 12(2):153--157, 1947.

\bibitem{mcp2025spec}
{Model Context Protocol Contributors}.
\newblock {Model Context Protocol} specification (revision 2025-06-18).
\newblock \url{https://modelcontextprotocol.io/specification/2025-06-18}, 2025.
\newblock Accessed: 2026-11-19.

\bibitem{monperrus2026osaudit}
Martin Monperrus.
\newblock {osaudit}: Automated open-science audit.
\newblock \url{https://github.com/ASSERT-KTH/osaudit}, 2026.
\newblock Accessed: 2026-11-19.

\bibitem{nabaos2026}
{NabaOS Contributors}.
\newblock {NabaOS}: Sandboxed agent runtime with {Ed25519}-signed
  constitutions.
\newblock \url{https://github.com/nabaos/nabaos}, 2026.
\newblock Accessed: 2026-11-19.

\bibitem{owasp2026agentic}
{OWASP GenAI Security Project}.
\newblock {OWASP} {Top} 10 for {Agentic} {Applications}.
\newblock \url{https://genai.owasp.org/}, 2026.
\newblock Accessed: 2026-11-19.

\bibitem{pipelock2026}
{Pipelab Contributors}.
\newblock {Pipelock}: Open-source {AI} agent firewall for {MCP} security and
  mediator-signed action receipts.
\newblock \url{https://github.com/pipelab-org/pipelock}, 2026.
\newblock Accessed: 2026-11-19.

\bibitem{rebedea2023nemo}
Traian Rebedea, Razvan Dinu, Makesh~Narsimhan Sreedhar, Christopher Parisien,
  and Jonathan Cohen.
\newblock {NeMo Guardrails}: A toolkit for controllable and safe {LLM}
  applications with programmable rails.
\newblock In {\em Proceedings of the 2023 Conference on Empirical Methods in
  Natural Language Processing: System Demonstrations}, pages 431--445.
  Association for Computational Linguistics, 2023.

\bibitem{wang2020minilm}
Wenhui Wang, Furu Wei, Li~Dong, Hangbo Bao, Nan Yang, and Ming Zhou.
\newblock {MiniLM}: Deep self-attention distillation for task-agnostic
  compression of pre-trained transformers.
\newblock In {\em Advances in Neural Information Processing Systems},
  volume~33, pages 5776--5788, 2020.

\bibitem{wilson1927probable}
Edwin~B. Wilson.
\newblock Probable inference, the law of succession, and statistical inference.
\newblock {\em Journal of the American Statistical Association},
  22(158):209--212, 1927.

\bibitem{zambo2026aer1}
Brennan Zambo.
\newblock {AER-1}: A portable execution receipt for {AI} agent tool calls.
\newblock IETF Internet-Draft \texttt{draft-zambo-aer1-26} (Work in Progress),
  \url{https://datatracker.ietf.org/doc/draft-zambo-aer1/26/}, 2026.
\newblock Accessed: 2026-11-20.

\end{thebibliography}
}

\end{document}